\documentclass[fleqn,usenatbib]{mnras}

\usepackage[dvipsnames]{xcolor}

\usepackage[T1]{fontenc}
\usepackage{ae,aecompl}

\usepackage{graphicx}	
\usepackage{amsmath}	
\usepackage{amssymb}	
\usepackage{lipsum}
\usepackage{rotating}
\usepackage{lscape}
\usepackage{float}
\usepackage{afterpage}

\usepackage{newtxtext,newtxmath}
\usepackage{appendix}

\def\gtorder{\mathrel{\raise.3ex\hbox{$\rangle $}\mkern-14mu
     \lower0.6ex\hbox{$\sim$}}}
\def\ltorder{\mathrel{\raise.3ex\hbox{$\langle $}\mkern-14mu
     \lower0.6ex\hbox{$\sim$}}}

\title[Dynamical Properties of Thin MADs]{Magnetic support, wind-driven accretion, coronal heating, and fast outflows in a thin magnetically arrested disc}

\author[N. Scepi et al.]{
Nicolas Scepi$^{1,2}$\thanks{E-mail: nicolas.scepi@gmail.com},
Mitchell C. Begelman$^{1,3}$
and Jason Dexter$^{1,3}$
\\
$^{1}$JILA, University of Colorado and National Institute of Standards and Technology, 440 UCB, Boulder, CO 80309-0440, USA \\
$^{2}$School of Physics and Astronomy, University of Southampton, Highfield, Southampton, SO17 1BJ, UK\\
$^{3}$Department of Astrophysial and Planetary Sciences, University of Colorado, 391 UCB, Boulder, CO 80309-0391, USA}

\date{Accepted XXX. Received YYY; in original form ZZZ}

\pubyear{2020}

\begin{document}
\label{firstpage}
\pagerange{\pageref{firstpage}--\pageref{lastpage}}
\maketitle
\begin{abstract}
 Accretion discs properties should deviate from standard theory when magnetic pressure exceeds the thermal pressure. To quantify these deviations, we present a systematic study of the dynamical properties of magnetically arrested discs (MADs), the most magnetized type of accretion disc. Using an artificial cooling function to regulate the gas temperature, we study MADs of three different thermal thicknesses, $h_\mathrm{th}/r=0.3, 0.1$ and $0.03$. We find that the radial structure of the disc is never mostly supported by the magnetic field. In fact, thin MADs are very near Keplerian. However, as discs gets colder, they become more magnetized and the largest deviations from standard theory appear in our thinnest disc with $h_\mathrm{th}/r=0.03$. In this case, the disc is much more extended vertically and much less dense than in standard theory because of vertical support from the turbulent magnetic pressure and wind-driven angular momentum transport that enhances the inflow speed. The thin disc also dissipates a lot of thermal energy outside of $z/r = \pm 0.03$ and a significant fraction of this dissipation happens in mildly relativistic winds. The enhanced dissipation in low-density regions could possibly feed coronae in X-ray binaries (XRBs) and active galactic nuclei (AGN). Wind-driven accretion will also impact the dynamical evolution of accretion discs and could provide a mechanism to explain the rapid evolution of changing-look AGN and the secular evolution of XRBs. Finally, our MAD winds have terminal velocities and mass loss rates in good agreement with the properties of ultra-fast outflows observed in AGN.
\end{abstract}



\makeatletter
\let\origsection\section
\renewcommand\section{\@ifstar{\starsection}{\nostarsection}}

\newcommand\nostarsection[1]
{\sectionprelude\origsection{#1}\sectionpostlude}

\newcommand\starsection[1]
{\sectionprelude\origsection*{#1}\sectionpostlude}

\newcommand\sectionprelude{%
  \vspace{-1mm}
}

\newcommand\sectionpostlude{%
  \vspace{-1mm}
}
\makeatother

\section{Introduction}
The thin disc model, the standard model of accretion disc theory \citep{Shakura,novikov1973}, has been generally successful in explaining the thermal UV radiation from active galactic nuclei (AGN), the soft X-ray spectrum of X-ray binaries (XRBs) and the spectrum of cataclysmic variables (CVs). However, the standard model alone cannot account for the rich phenomenology observed in accretion discs around compact objects. For example, XRBs and changing-look AGN evolve on short time scales that require very efficient angular momentum transport compared to the traditional turbulent transport invoked in standard theory \citep{tetarenko2018,lawrence2018,dexter2019}. XRBs and AGN also exhibit hard X-ray spectra that cannot be accounted for by standard theory \citep{elvis1994,remillard2006}. This hard emission is usually referred to as coronal emission since it must originate from hot, optically thin gas \citep{fabian2015}. All types of accretion discs show the presence of outflows in the form of low-density, high-velocity collimated jets \citep{burrows1996,fender1997,mirabel1999,corbel2000,merloni2003,lister2009} or slower, denser and more extended winds \citep{cordova1982,crenshaw2003,ponti2012,louvet2018} that are not accounted for in standard theory. Finally, the standard model fails to explain the emission of black holes accreting near the Eddington limit where the standard model is thermally unstable \citep{lightman1974,shakura1976,piran1978}.

Over the years, several authors have suggested that strong magnetic fields in accretion discs could resolve these outstanding issues of accretion theory. More specifically, it has been proposed that strong magnetic pressure could support the vertical structure of the disc to provide faster inflow \citep{gaburov2012,dexter2019} and thermal stability \citep{begelman2007,sadowski2016}. Additional large-scale magnetic  torques, such as provided by magnetic outflows \citep{blandford1982,ferreira1993,scepi2018b}, could also provide faster inflow and produce shorter evolution time scales \citep{tetarenko2018,scepi2021}. A larger inflow speed would also reduce the density in the disc and could produce the hard X-ray emission in XRBs if the disc becomes optically thin \citep{ferreira2006,petrucci2010,marcel2018a}. Finally, magnetic fields could lead to strong magnetic dissipation outside the dense core of the disc and so provide additional energy to produce the optically thin, hard X-ray emission \citep{liang1977,field1993,miller2000,merloni2001,jiang2019}.

All of these effects should be maximized when the magnetic field is strongest. The most magnetized type of disc around an accreting black hole is called a magnetically arrested disc (MAD) \citep{igumenshchev2008}. In MADs, the magnetic flux on the black hole and in the disc has reached saturation \citep{tchekhovskoy2011,begelman2022}. Despite their name, the accretion in MADs is never arrested \citep{begelman2022} although the magnetic field does perturb the accretion \citep{igumenshchev2008,mckinney2012}, which proceeds through non-axisymetric features in the disc. In hot, puffy MADs, which are the preferred models to explain observations from lower-luminosity black holes \citep{dexter2019,jimenez2020,eht2021,scepi2021,dexter2021}, the saturation of the flux on the black hole is believed to happen through intermittent large-scale reconnection events between the two polarities of a strong jet originating from the black hole \citep{ripperda2022}. These reconnection events eject the magnetic flux away from the black hole before it is again advected by the accretion flow. The mechanism that regulates the magnetic flux in the disc further away from the black hole is less clear, although it has been proposed that in hot MADs the regulation happens through convective/interchange instabilities involving the toroidal field \citep{begelman2022}. 

The effect of the magnetic field in cold, thin MADs, which should be better models of luminous systems, remains largely unexplored because of the high numerical resolution required to study these discs. However, it is in the thin disc regime that we should naively expect the impact of the magnetic field to be the greatest  since the thermal energy is smaller there. In fact, the few existing studies of thin MADs show large deviations from standard theory regarding their radiative efficiency, inflow speed, thermal stability and/or equilibrium \citep{avara2016,sadowski2016,morales2018,liska2019,liska2022}. This motivates a systematic study of how the effects of magnetic field change when going from thick to thin discs, which is the purpose of this paper. 

In this paper, we show the results of a series of three MAD simulations ranging from hot, thick discs to cold, thin discs. For each simulation, we show how the magnetic field affects the radial and vertical structure of the disc and the role it plays in angular momentum transport, energy dissipation and energy extraction through ejection of matter. We then discuss our results in the broad context of accretion discs around compact objects to show how strong magnetic fields can alleviate the limitations of standard theory. 

\section{Methods}\label{sec:methods}

\subsection{Basic equations}

Throughout this paper, we use Greek letters to express sums over time and spatial components and Latin letters to express sums over spatial components only. We use units where $G=c=M=1$ and a factor of $\sqrt{4\pi}$ is implicitly included in the electromagnetic field. We use Kerr-Schild coordinates and our metric convention, $(-,+,+,+)$, is the same as in \cite{misner1973}.

We use the public code {\sc Athena++} \citep{white2016,stone2020} that solves the general relativistic equations of ideal magnetohydrodynamics (GRMHD) written in the following form:
\begin{gather}
    \partial_t(\sqrt{-g}\rho u^t)+\partial_j(\sqrt{-g}\rho u^j)=0, \label{eq:continuity} \\
    \partial_t(\sqrt{-g} T^t_\mu) + \partial_j(\sqrt{-g} T^j_\mu) =\sqrt{-g}\Gamma^\sigma_{\lambda \mu}T^{\lambda_\sigma}-\sqrt{-g}\mathcal{F}_\mu, \label{eq:stress_energy_tensor} \\
    \nabla_\mu \text{*}F^{\mu\nu} =0, \label{eq:induction}
\end{gather}
where $\rho$ is the fluid frame density, $u^\mu$ is the coordinate frame 4-velocity, $g$ is the metric determinant and $\Gamma^\sigma_{\lambda \mu}$ is the Christoffel symbol. The last term in \autoref{eq:stress_energy_tensor} is a source term describing the artificial cooling function that we will present in \S\ref{sec:cooling}. We also define the 3-vector velocity $V^i=u^i/u^t$. Here $T^{\mu\nu}$ and $\text{*}F^{\mu\nu}$ are the ideal MHD stress-energy tensor and the dual electromagnetic tensor, respectively, defined as 
\begin{gather}
    T^{\mu\nu} = (\rho h +b_\lambda b^\lambda)u^\mu u^\nu + (p_\mathrm{gas}+p_\mathrm{mag})g^{\mu\nu}-b^\mu b^\nu,\\
    \text{*}F^{\mu\nu} = b^\mu u^\nu - b^\nu u^\mu,
\end{gather}
where $p_\mathrm{gas}$ is the gas pressure, $h=1+\gamma_\mathrm{ad} p_\mathrm{gas}/(\gamma_\mathrm{ad}-1)$ is the enthalpy per unit mass, $\gamma_\mathrm{ad}=5/3$ is the adiabatic index, $b^\mu$ is the fluid frame 4-magnetic field, $p_\mathrm{mag}=b_\lambda b^\lambda /2$ and $g^{\mu\nu}$ is the metric tensor. To solve \autoref{eq:continuity}, \ref{eq:stress_energy_tensor} and \ref{eq:induction}, we use a second-order van Leer integrator scheme in time combined with a HLLE Riemann solver \citep{einfeldt1988} and a piece-wise linear spatial reconstruction method \citep{vanleer1974}.

\subsection{Numerical set-up}

We initialize a \cite{fishbone1976} torus  around a spinning black hole of spin $a/M=0.9375$. We set the inner boundary of the torus at 16.45 $r_g$ and the pressure maximum at 34 $r_g$, where $r_g\equiv GM/c^2$ is the gravitational radii with $M$ being the mass of the black hole, $G$ the gravitational constant and $c$ the speed of light. The simulated domain is defined by $r\in[1.125,1500]\:r_g$, $\theta\in[0,\pi]$ and $\phi\in[0,2\pi]$. The inner boundary is chosen so that there are at least 8 cells inside the event horizon. We use up to four levels of static mesh refinement in the disc, allowing for an effective resolution of $1024\times512\times1024$ in our most resolved cases. For our two thickest simulations we use four refinement levels to concentrate the resolution in the disc while still having a relatively high resolution in the wind and jet regions.  For our thinnest simulation we use four refinement levels to concentrate the resolution only in the disc while having less resolution in the wind and jet. The refinement levels are described in the left columns of \autoref{table:refinement}. We also show the extra-refinement level that we used for testing the convergence of our simulation with $h_\mathrm{th}/r=0.03$. Since we do not find any quantitative difference between our higher effective resolution ($2048\times1024\times2048$) simulation and our lower effective resolution ($1024\times512\times1024$) simulation (see \S\ref{sec:convergence} and \S\ref{sec:appendixA}), we only report the results for the lowest resolution simulation, which is easier to analyze.
 
\begin{table}
\centering
\begin{tabular}{||c c c c c c||} 
 \hline
 Level & r-range & $\theta$-range & $N_{r,\mathrm{eff}}$ & $N_{\theta,\mathrm{eff}}$ & $N_{\phi,\mathrm{eff}}$\\ [0.5ex] 
 \hline
 \multicolumn{6}{||c||}{$h_\mathrm{th}/r=0.3$ and $0.1$}\\
 \hline
  0 & 1.125-1500 & 0-$\pi$ & 64 & 32 & 64 \\
  1 & 10-240 & 0-$\pi$ & 128 & 64 & 128 \\
  1 & 1.125-240 & $\pi/8$-$7\pi/8$ & 128 & 64 & 128 \\
  2 & 1.125-100 & $3\pi/16$-$13\pi/16$ & 256 & 128 & 256 \\
  3 & 1.125-60 & $5\pi/16$-$11\pi/16$ & 512 & 256 & 512 \\ 
 \hline
  \multicolumn{6}{||c||}{$h_\mathrm{th}/r=0.03$ }\\
  \hline
  0 & 1.125-1500 & 0-$\pi$ & 64 & 32 & 64 \\
  1 & 1.125-240 & $\pi/4$-$3\pi/4$ & 128 & 64 & 128 \\
  2 & 1.125-100 & $3\pi/8$-$5\pi/8$ & 256 & 128 & 256 \\
  3 & 1.125-60 & $7\pi/16$-$9\pi/16$ & 512 & 256 & 512 \\ 
  4 & 1.125-40 & $15\pi/32$-$17\pi/32$ & 1024 & 512 & 1024 \\ 
  (5) & (1.125-20) & ($31\pi/64$-$33\pi/64$) & (2048 & 1024 & 2048) \\ 
 \hline

\end{tabular}
\caption{Description of the static mesh refinement levels for our three simulations. The $\phi$-range is always $2\pi$. The last line with numbers in parenthesis shows the additional level of refinement we used for our convergence test in \S\ref{sec:convergence}.}
\label{table:refinement}
\end{table}

To stabilize the numerical scheme, we use floors for the density and thermal pressure  of  $10^{-8}$ and $10^{-10}$, respectively, in code units where the initial maximum values for the density and the thermal pressure in the torus are 1 and 0.005. The plasma $\beta$-parameter, $\beta\equiv p_\mathrm{gas}/p_\mathrm{mag}$, has a floor of 0.01, and we impose ceilings for the fluid Lorentz factor, $\gamma$, and the rest-mass magnetization, $\sigma\equiv 2 p_\mathrm{mag}/\rho$, of 50 and 100, respectively. 

\subsection{Cooling function}\label{sec:cooling}
We use an artificial cooling function implemented as in \cite{noble2009} to control the gas temperature. The cooling function is implemented as a source term, $\mathcal{F}_\mu$, in \autoref{eq:stress_energy_tensor} where $\mathcal{F}_\mu$ represents the amount of radiated energy-momentum per unit of volume and time in the coordinate frame.

The cooling function targets a radius-dependent isothermal sound speed $c_\mathrm{s}$ corresponding to a constant thermal aspect ratio of the disc, 
\begin{equation}
    \frac{h_\mathrm{th}}{r}\equiv \sqrt{\frac{2}{\pi}}\frac{c_\mathrm{s}}{v_\mathrm{K}},
\end{equation}
where  $v_\mathrm{K}$ is the relativistic Keplerian velocity around a Kerr black hole. As in \cite{noble2009}, we define $v_\mathrm{K}= R/(R^{3/2}+a/M)$ for $R>R_\mathrm{ISCO}$ and as the orbital velocity of a particle with the specific energy and angular momentum of the circular orbit at the ISCO for $R<R_\mathrm{ISCO}$. Here $R$ is the cylindrical radius. We set the characteristic cooling time scale to $\mathrm{max}(0.125 t_\mathrm{orb},\Delta t)$ where $t_\mathrm{orb}=R/v_\mathrm{K}$ and $\Delta t$ is the timestep of the simulation.

This cooling function offers a very easy way to track energy dissipation in the simulation \citep{noble2009}.  However, we note that the cooling function used in \cite{noble2009} only cools the bound material, i.e., the bulk of the disc but neither the wind nor the jet. In order to track the energy dissipation in regions that are more extended vertically than the disc (see \S\ref{sec:heating}), we therefore modified the cooling function to cool all regions with $\sigma<1$; this avoids cooling the jet while still cooling the unbound wind region. We also ran a second version of each simulation, cooling only the bound regions, to compare our results with the literature. Except in \autoref{fig:efficiencies}, we only show the results of the simulations where both the disc and wind are cooled. We find that this distinction between cool and hot wind makes a difference only when considering the efficiencies of the outflows (especially for the hot disc with $h_\mathrm{th}/r=0.3$) but does not affect the rest of the conclusions drawn in this paper.

We use this cooling function to explore a parameter space going from hot to cold MADs with targeted thermal heights of $h_\mathrm{th}/r=0.3, 0.1$ and $ 0.03$ (see top panel of \autoref{fig:h_rot}). We emphasize that the thermal height does not necessarily reflect the geometrical height of the disc because of magnetic support, as we will see in \S\ref{sec:vertical}.

For our simulations with $h_\mathrm{th}/r=0.3, 0.1$ we cool the initial torus right from the beginning of our simulation. However, for $h_\mathrm{th}=0.03$ we start by cooling the initial torus to $h_\mathrm{th}/r=0.3$ for 20,000 $r_g/c$, let the disc become MAD, and then cool the disc to $h_\mathrm{th}/r=0.03$. We find that this two-step procedure prevents a violent transient in which hot material from the initial torus is suddenly cooled at very large radii and arrives on the inner disc in an intermittent and non-symmetric way with respect to the midplane. 

\subsection{Initial magnetic field}
The initial torus is threaded by net vertical magnetic flux with an initial magnetization of $\beta_\mathrm{torus}\approx10^2$, where $\beta_\mathrm{torus}$ is the ratio of thermal to magnetic pressure at the midplane in the intial torus. We use the following magnetic potential to initialize the magnetic field
\begin{equation}
    A_\phi(R) = \begin{cases}
\mathcal{C}_\beta\tanh((R_\mathrm{in}-R_0)/w) &\text{if $R <  R_\mathrm{in}$}\\
\mathcal{C}_\beta\tanh((R-R_0)/w) &\text{if $R_\mathrm{in} <  R <  R_\mathrm{out}$}\\
\mathcal{C}_\beta\tanh((R_\mathrm{out}-R_0)/w) &\text{if $R >  R_\mathrm{out}$},
\end{cases} 
\end{equation}
where $R$ is the cylindrical radius. This magnetic potential gives an almost flat profile of $\beta_\mathrm{torus}$ as a function of the cylindrical radius in the torus and allows the field to go smoothly to 0 at $R_\mathrm{in}$ and $R_\mathrm{out}$ to better preserve the solenoidal condition than in the case of a purely constant vertical magnetic field. We use $R_\mathrm{in}=16.45\:r_g$, $R_0=34\:r_g$, $R_\mathrm{out}=70\:r_g$ and $w=30\:r_g$. The normalization constant is set by $\beta_\mathrm{torus}$.\\

\section{Time-dependent evolution of MADs}\label{sec:time_dependent}

\begin{figure*}
\includegraphics[trim={0 10mm 0 0},width=\textwidth]{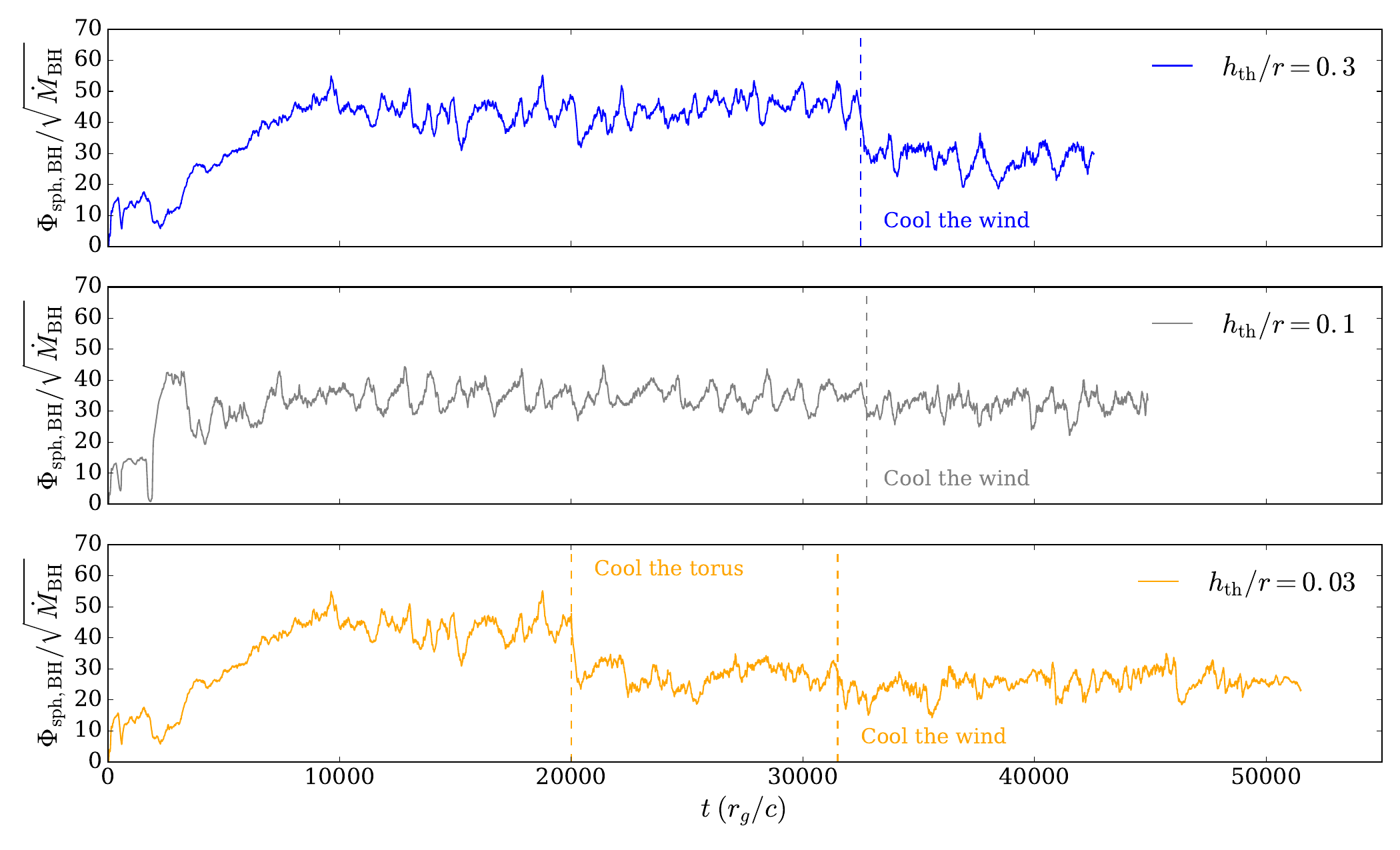}
\caption{$\Phi_\mathrm{sph,BH}/\sqrt{\dot{M}_\mathrm{BH}}$ as a function of time for all of our simulations (from top to bottom). We show the times at which we start cooling the wind, which are $32500,\:32750$ and $31500\:r_g/c$ for $h_\mathrm{th}/r=0.3,0.1$ and $0.03$, respectively. For the simulation with $h_\mathrm{th}/r=0.03$, we also show the time at which we start to cool the torus to a thin disc (at $20,000\:r_g/c$). We find that all of our simulations saturate to a constant value of $\Phi_\mathrm{sph,BH}/\sqrt{\dot{M}_\mathrm{BH}}$. However, the value at saturation depends on both the targeted $h_\mathrm{th}/r$ and on whether we cool the wind or not.}
\label{fig:time_phi}
\end{figure*}

MADs are well-known for their magnetic flux saturation on the black hole \citep{tchekhovskoy2011}. The saturation operates through cycles of flux eruptions and flux accumulation once the normalized magnetic flux (the spherical magnetic flux divided by the square root of the accretion rate) has reached a certain threshold that depend on the disc thickness and the black hole spin \citep{tchekhovskoy2011,mckinney2012,avara2016,liska2022,narayan2022}. We show in \autoref{fig:time_phi} the time evolution of the normalized spherical magnetic flux, $\Phi_\mathrm{sph,BH}/\sqrt{\dot{M}_\mathrm{BH}}$, as in \cite{tchekhovskoy2011}, where the spherical magnetic flux is defined as
\begin{equation}
    \Phi_\mathrm{sph}(r) = \int_{\mathcal{H}(r)}\sqrt{4\pi} B^r \sqrt{-g}d\theta d\phi,
\end{equation}
with $\mathcal{H}(r)$ the surface of one hemisphere of radius $r$. We see that $\Phi_\mathrm{sph,BH}/\sqrt{\dot{M}_\mathrm{BH}}$ saturates to a constant value for all of our simulations.  As already noted, we find that the value of saturation depends on $h_\mathrm{th}/r$. Indeed, before the wind was cooled, $\Phi_\mathrm{sph,BH}/\sqrt{\dot{M}_\mathrm{BH}}$ was $\approx45, 36$ and $28$ for $h_\mathrm{th}/r=0.3, 0.1$ and $0.03$ respectively. The saturation value for our hottest MAD is quite similar to the value of $\approx50$ usually reported in the literature for thick MADs with highly spinning black holes \citep{tchekhovskoy2011,mckinney2012}. The value for $h_\mathrm{th}/r=0.1$ is higher than the value of $\approx25$ reported in \cite{avara2016} but we note that they used a lower spin than us ($a=0.5$) for their simulation. Finally, the simulation of \cite{liska2022} reports a slightly larger saturation value than us ($\approx30-35$ compared to $\approx28$) for their simulation with $h_\mathrm{th}/r=0.03$.

Interestingly, we find that the saturation value of $\Phi_\mathrm{sph,BH}/\sqrt{\dot{M}_\mathrm{BH}}$ does not depend only on $h_\mathrm{th}/r$ in the disc but also on whether we cool the wind or not. We see that as soon as we cool the wind the value of $\Phi_\mathrm{sph,BH}/\sqrt{\dot{M}_\mathrm{BH}}$ goes down to $\approx28, 34$ and $25$ for $h_\mathrm{th}/r=0.3, 0.1$ and $0.03$, respectively. In \S\ref{sec:winds}, we will see that the power of the wind also decreases when the wind is cooled and that this effect is more dramatic for $h_\mathrm{th}/r=0.3$. These observations seem to imply that the saturation value of $\Phi_\mathrm{sph,BH}/\sqrt{\dot{M}_\mathrm{BH}}$ is related to the amount of pressure that the wind can exert against the magnetic field lines anchored on the BH, i.e. the jet. This would be in line with a scenario in which the level of magnetic flux saturation is determined by a balance between the wind+accretion disc pressure and the jet pressure. The higher the wind+accretion disc pressure, the higher the power attained by the jet before it forces the opposing fields on each side of the disc to reconnect   \citep{ripperda2022}, effectively regulating the magnetic flux on the black hole.

\begin{figure}
\includegraphics[trim={0 10mm 0 0},width=90mm]{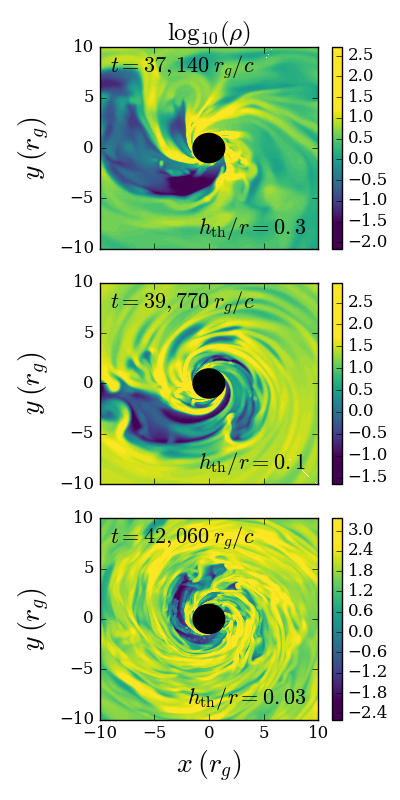}
\caption{Snapshots of the midplane density during a magnetic flux eruption event for $h_\mathrm{th}/r=0.3,0.1$ and 0.03 are shown in the top, middle and bottom panels, respectively. We see the same structure, characteristic of the MAD state, with low density cavities appearing for all values of  $h_\mathrm{th}/r$.}
\label{fig:rho_eruption}
\end{figure}

As mentioned above, a picture is now emerging in which the flux eruptions are caused by reconnection events in the midplane of the disc near the black hole \citep{ripperda2022}. These events create low density, high magnetization, slowly rotating cavities that are expelled from the black hole and propagate to large radii in the disc \citep{porth2021}. These flux eruptions and in particular these cavities are regarded as distinctive features of MADs and have attracted interest because of their potential to accelerate non-thermal electrons \citep{scepi2022,ripperda2022}.  We plot on \autoref{fig:rho_eruption} a snapshot of the density in the midplane of all of our simulations during a flaring event. We see that all of our simulations, including our thinnest simulation with $h_\mathrm{th}/r=0.03$, show the same behavior as reported previously in the literature with the creation of low-density, high magnetization cavities in the disc. This makes us confident that all of our simulations are indeed in a MAD state in which the flux has saturated and that the different saturation levels are related to different disc+wind properties as explained above. Note that the presence of similar low-density, high magnetization cavities in both thin and thick discs suggests that  non-thermal particles can be accelerated in high luminosity discs near the black hole \citep{hankla2022} in a similar way to what has been proposed for low-luminosity discs \citep{scepi2022,ripperda2022}.

\section{Time-averaged structure of MADs}\label{sec:time_averaged}

The results shown from this point in the paper are obtained by time- and azimuthally averaging, noted as $\langle \rangle_{t,\phi}$ the outputs of our simulations. The data are time-averaged between $34,000$ and $42,000\:r_g/c$, $35,000$ and $42,000\:r_g/c$ and $40,000$ and $44,000\:r_g/c$ for $h_\mathrm{th}/r=0.3,0.1$ and $0.03$ respectively. Note that the time averaging windows are chosen so that the disc is in quasi-steady state within $10\:r_g$. In what follows, we refer to the laminar component of $X\times Y$ as $\langle X\rangle_{t,\phi} \langle Y\rangle_{t,\phi}$ and the turbulent component as $\langle X\times Y \rangle_{t,\phi} - \langle X\rangle_{t,\phi} \langle Y\rangle_{t,\phi}$. From now on all quantities are by default averaged quantities and we will not use the symbol $\langle\rangle_{t,\phi}$ anymore.

\subsection{Radial structure}

\begin{figure}
\includegraphics[trim={0 10mm 0 0},width=90mm]{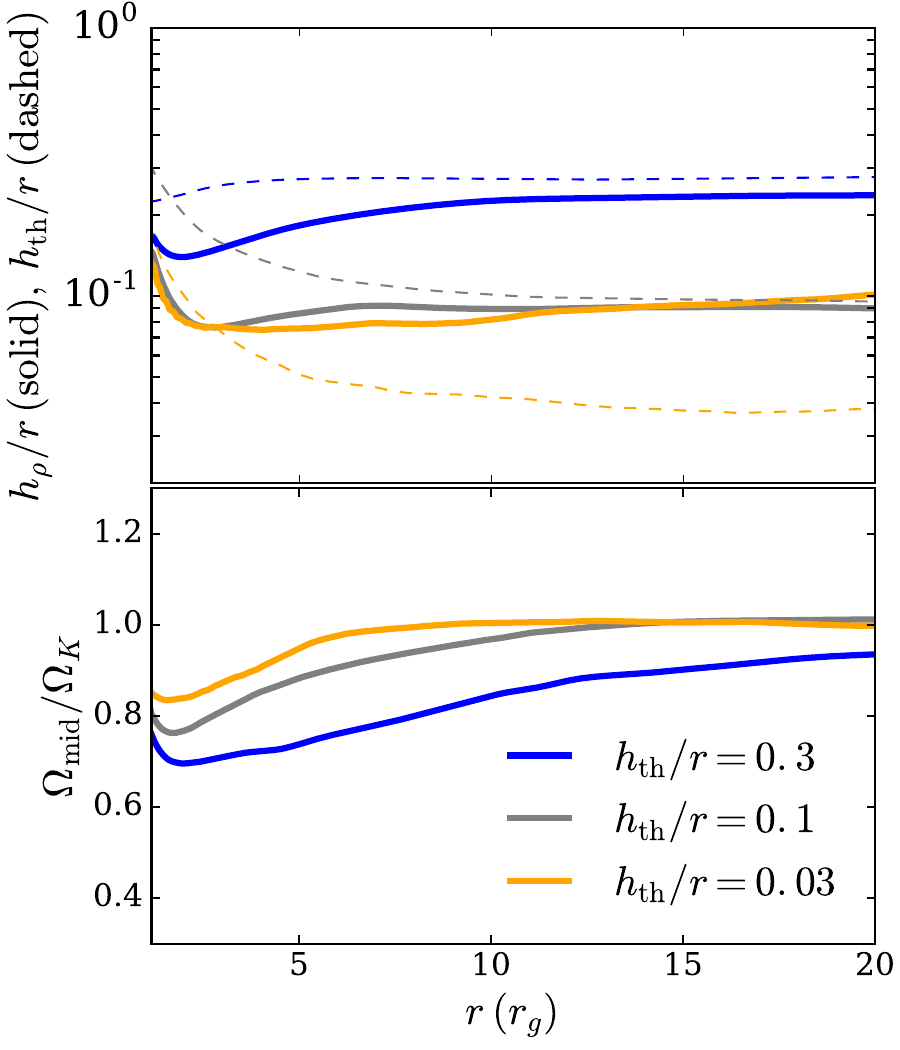}
\caption{Top panel: Density scale height (solid line) and thermal scale height (dashed) as a function of radius for our MAD simulations.  Bottom panel: Rotation rate in the midplane divided by the Keplerian rotation rate as a function of radius. The thin disc is  very close to Keplerian but has a much larger density scale height than its thermal scale height.}
\label{fig:h_rot}
\end{figure}

\begin{figure}
\includegraphics[trim={0 10mm 0 0},width=90mm]{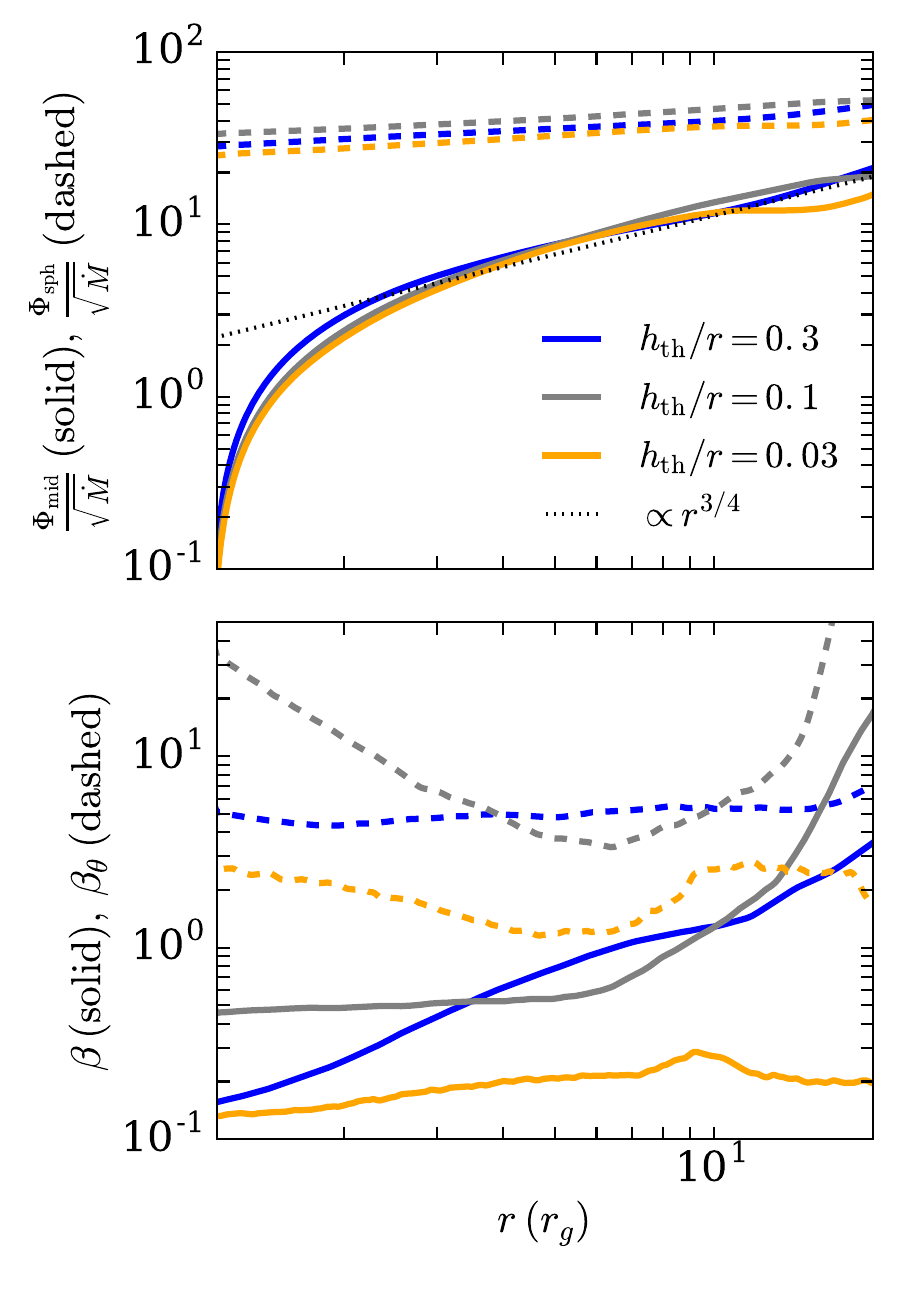}
\caption{ Top panel: Magnetic flux in the midplane (solid lines) and spherical magnetic flux (dashed lines) as a function of radius. Bottom panel: Plasma $\beta$ parameter in the midplane as a function of radius for all of our simulations. The dashed and solid lines respectively show $\beta$ computed using the magnetic pressure from the latitudinal magnetic field only and $\beta$ computed using the total magnetic pressure. The saturated magnetic flux on the black hole varies with $h_\mathrm{th}/r$ and is minimum for $h_\mathrm{th}/r=0.03$ but the saturated midplane magnetic flux is the same in all simulations. Consequently, the thinner the disc the more magnetized it is.}
\label{fig:phi_beta}
\end{figure}

In \cite{begelman2022}, two criteria were used to determine the radius up to which a hot disc was in the MAD state: the sub-Keplerian rotation profile and the saturation of the magnetic flux in the disc. On the bottom panel of \autoref{fig:h_rot}, we plot the rotation rate in the midplane, defined as
\begin{equation}
    \Omega_\mathrm{mid} = (u^\phi/u^t)_\mathrm{mid},
\end{equation}
normalized to the Keplerian rotation rate, and on the first panel of \autoref{fig:phi_beta} we plot the spherical and midplane magnetic flux normalized by the square root of the accretion rate as a function of radius, where the midplane magnetic flux is defined as 
\begin{equation}
    \Phi_\mathrm{mid}(r) = \int_{r,\phi, \theta=\pi/2}\sqrt{4\pi}  B^\theta \sqrt{-g} dr d\phi.
\end{equation}

\cite{begelman2022} found that in a hot, adiabatic MAD, where no cooling function was used and the thermal scale height was $\approx h_\mathrm{th}/r\approx0.4$, the rotation profile was $\approx0.5-0.6$ Keplerian. This was due to the radial gravity being mainly compensated for by, first, the gradient of thermal pressure, and second, the centrifugal force. We see from the second panel of \autoref{fig:h_rot} that the rotation profile becomes closer to Keplerian as the MAD disc becomes colder with a profile that is around $0.7-0.8$ Keplerian for $h_\mathrm{th}/r=0.3$ and $0.9-1$ for $h_\mathrm{th}/r=0.03$ for regions with $r\lesssim 10r_g$, which are the regions in inflow equilibrium. This faster when colder behaviour is due to the support of thermal pressure becoming smaller as the disc gets colder. We note that the magnetic pressure plays little role in the radial equilibrium of the disc.

The top panel of \autoref{fig:phi_beta} shows that the normalized magnetic flux in the midplane saturates to a constant value and profile independent of the thermal state of the disc. The profile goes as $\propto r^{3/4}$, the expected radial profile in a self-similar disc. We note that a profile going as $\propto r$ was reported in \cite{begelman2022}. It is unclear whether this difference is due to a different initial profile for the magnetic field or the use of a different equation of state.

We plot on the bottom panel of \autoref{fig:phi_beta} the radial profile of $\beta$ and $\beta_\theta$ for our three MADs. We see that as the disc becomes thinner it also becomes more magnetized. This is consistent with the fact that the normalized magnetic flux in the midplane stays the same for all MADs independently of the thermal scale height. Consequently, $\beta_\theta$ is larger by a factor of $\approx10$ for $h_\mathrm{th}/r=0.3$ than for $h_\mathrm{th}/r=0.03$. We see that $\beta$, which is computed on the total magnetic field and so depends mainly on the MRI-generated toroidal field, does not exactly follow that trend. Indeed, the value of $\beta$ drops near the black hole when $h_\mathrm{th}/r$ is large. This is due to the radial field of the jet dominating near the black hole when the disc is hot and puffy. However, it is interesting to note that, as the contribution from the jet becomes weaker, our discs approach a constant radial profile of $\beta$ as well as a constant value of $\beta_\theta$.

\begin{figure*}
\includegraphics[trim={0 10mm 0 0},width=\textwidth]{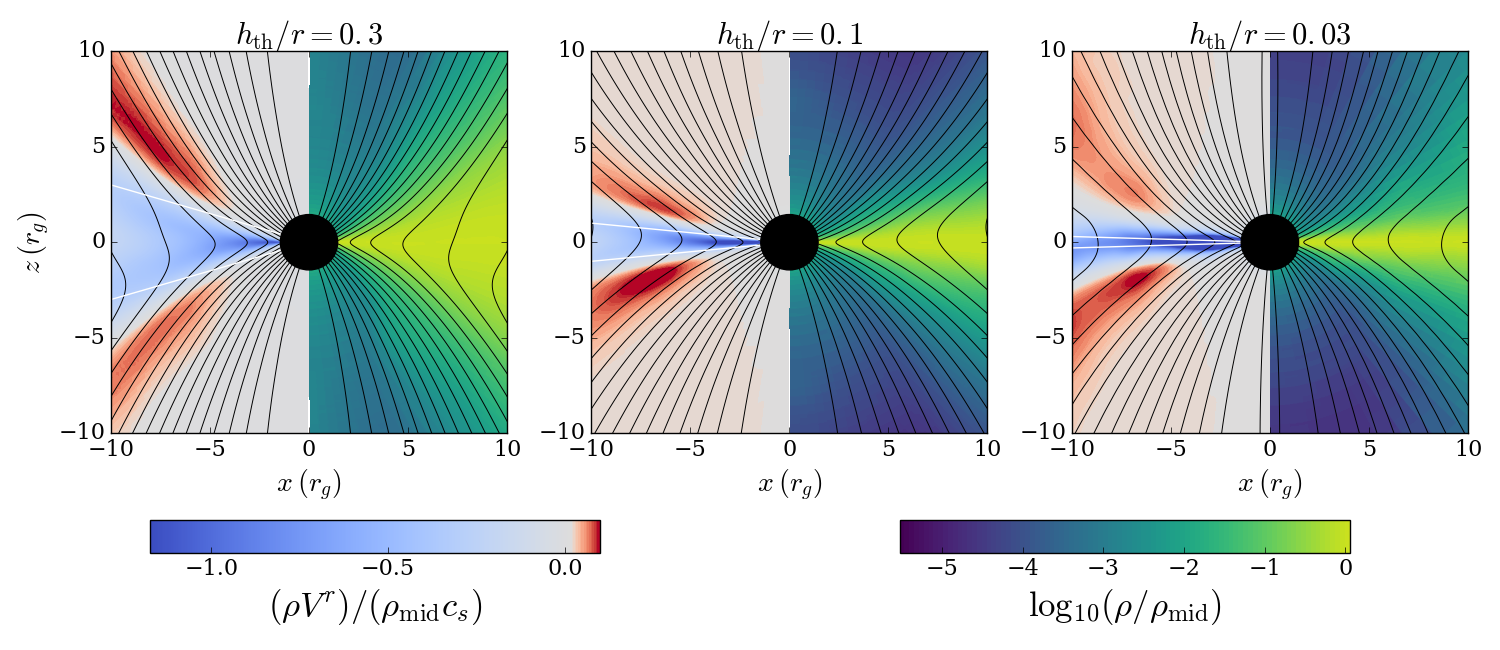}
\caption{Color maps of the density normalized to the midplane density (right panels) and the radial mass flux normalized to the midplane density times the sound speed (to see when the inflow becomes transonic) in the disc (left panels). The color map on the left panels is saturated in the blue to better show the dependence of the wind velocity with $h_\mathrm{th}/r$. The black lines represent the magnetic field lines in all panels. The white lines show the thermal scale height of the disc. We see that the density structure of the disc barely changes between $h_\mathrm{th}/r=0.1$ and $h_\mathrm{th}/r=0.03$ because of magnetic pressure support.}
\label{fig:rho_ur_mag}
\end{figure*}

\subsection{Vertical structure}\label{sec:vertical}

\autoref{fig:rho_ur_mag} shows maps of the density and radial velocity weighted by the density with poloidal magnetic field lines superimposed on both plots for our three MAD simulations. We see that as we go from hot to cold discs the density settles to a thinner disc. However, we can also see that the density structure from $h_\mathrm{th}/r=0.1$ to $h_\mathrm{th}/r=0.03$ barely changes. This can be seen more clearly in the top panel of \autoref{fig:h_rot} where we plot thermal scale heights as dashed lines and density scale heights as solid lines, as a function of radius. We see that for the MAD simulation with $h_\mathrm{th}/r=0.1$ the density scale height and the thermal scale height are close to one another. For $h_\mathrm{th}/r=0.3$, the density scale height is close to but slightly smaller than the thermal scale height. However, for the simulation with $h_\mathrm{th}/r=0.03$ the density scale height is three times larger than the thermal scale height, indicating a large deviation from standard theory.

\begin{figure*}
\includegraphics[trim={0 10mm 0 0},width=1.\textwidth]{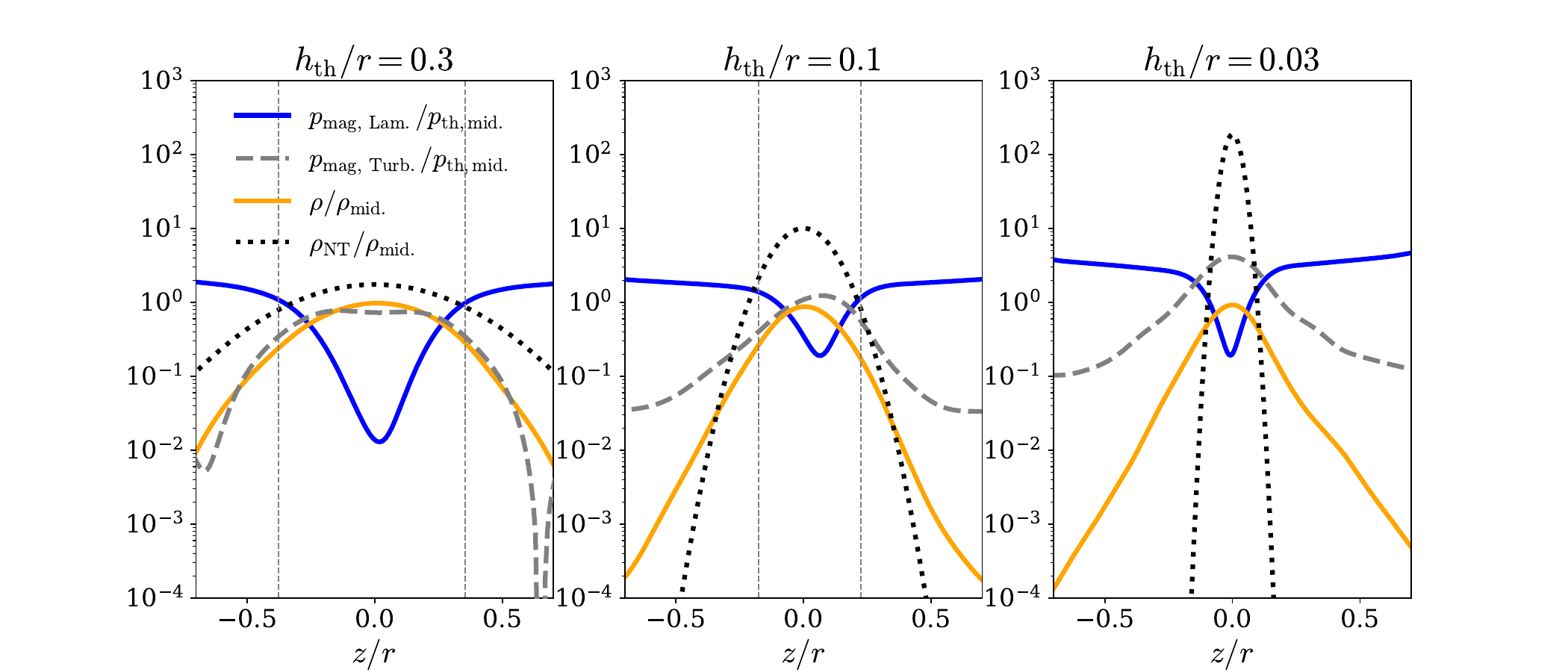}
\caption{The orange solid lines show the density as a function of $z/r$, normalized to the midplane density, that we measure in our simulations at $r=7\:r_g$. The black dotted lines show the expected density profile for a disc vertically supported by the gradient of thermal pressure (where the thermal pressure is given by our cooling function) in the standard, Novikov --Thorne (NT) theory. The expected density profile is normalized so as to give the same accretion rate as in our simulation and we fixed $\alpha=1$ for the theoretical profile. The blue solid line shows the laminar magnetic pressure and the dashed grey line shows the turbulent magnetic pressure. The left, middle and right panels show the results for $h_\mathrm{th}/r=0.3, 0.1$ and $0.03$, respectively. The grey dashed vertical lines show the surface of the disc defined as where the flow crosses the slow magneto-sonic point. The turbulent magnetic pressure is puffing up the $h_\mathrm{th}/r=0.03$ simulation to a much larger scale height than expected from thermal pressure. The  midplane density in our simulation is much lower than expected for a standard disc with the same gas temperature and accretion rate. }
\label{fig:pressure_dec}
\end{figure*}

The difference in the density profiles can also be seen in \autoref{fig:pressure_dec}, where we plot as a solid orange line the latitudinal profiles of the density in our simulations normalized to the midplane density. We also plot as a black dotted line the expected density profile if the disc was supported by the gradient of thermal pressure as in standard theory. The expected density profile is normalized so as to give the same accretion rate as in our simulation. Note that the density profiles in our simulations can differ from theory because of two effects: different effective scale heights or different accretion velocities. Indeed, the accretion rate is $\dot{M}\approx 4\pi R \rho H v_R$, where $R$ is the cylindrical radius, $H$ is the height of the disc and $v_R$ is the accretion velocity. A higher scale height and/or a higher velocity than expected would both give a lower density. In standard theory the effective scale height, $H$, is the same as $h_\mathrm{th}$ and the radial velocity is $v_R=\alpha(H/R)^2 v_\mathrm{K}$. We take here $\alpha=1$ for better comparison with our simulations (see \autoref{fig:stresses}) but note that standard models of accretion discs usually take lower values of $\alpha\approx 0.1$, which would accentuate even more the difference between standard models and our simulations. 

\begin{figure*}
\includegraphics[trim={0 10mm 0 0},width=1.\textwidth]{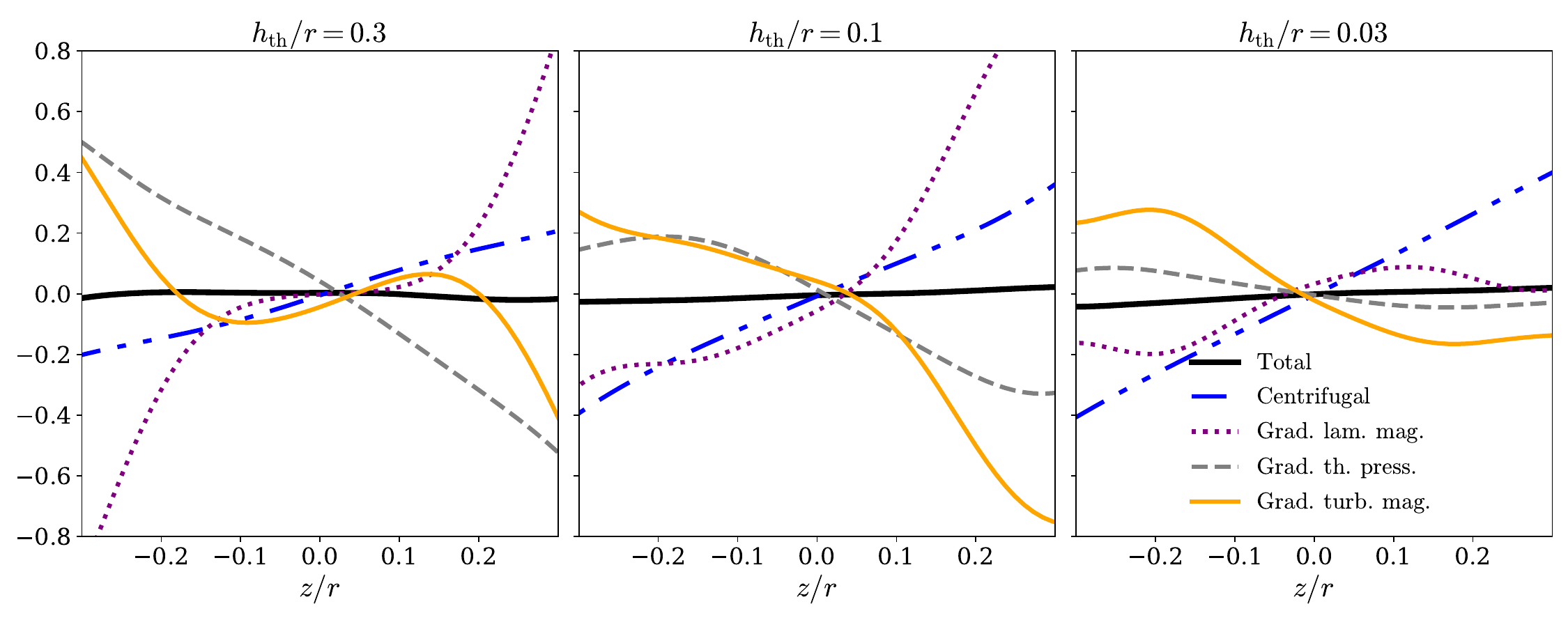}
\caption{Latitudinal force budget as a function of $z/r$ with  the results for our $h_\mathrm{th}/r=0.3, 0.1$ and $0.03$ simulations shown on the left, middle, and right panels, respectively. The blue dash-dotted line, the purple dotted line, the grey dashed line and the solid orange line show the contribution of the centrifugal force, the gradient of laminar magnetic pressure, the gradient of thermal pressure and the gradient of turbulent magnetic pressure, respectively, in the latitudinal equilibrium. The black line shows the sum of all forces, including the magnetic tension, the hoop stress and the poloidal acceleration forces that we do not show on this graph because of their secondary role in the latitudinal equilibrium. The force budget is measured at $r=7\:r_g$. When the disc is in latitudinal equilibrium the solid black line should go to zero. For $h_\mathrm{th}/r=0.03$, the turbulent magnetic pressure becomes the main force supporting the latitudinal structure.}
\label{fig:force_theta}
\end{figure*}

\begin{figure*}
\includegraphics[trim={0 10mm 0 0},width=1.\textwidth]{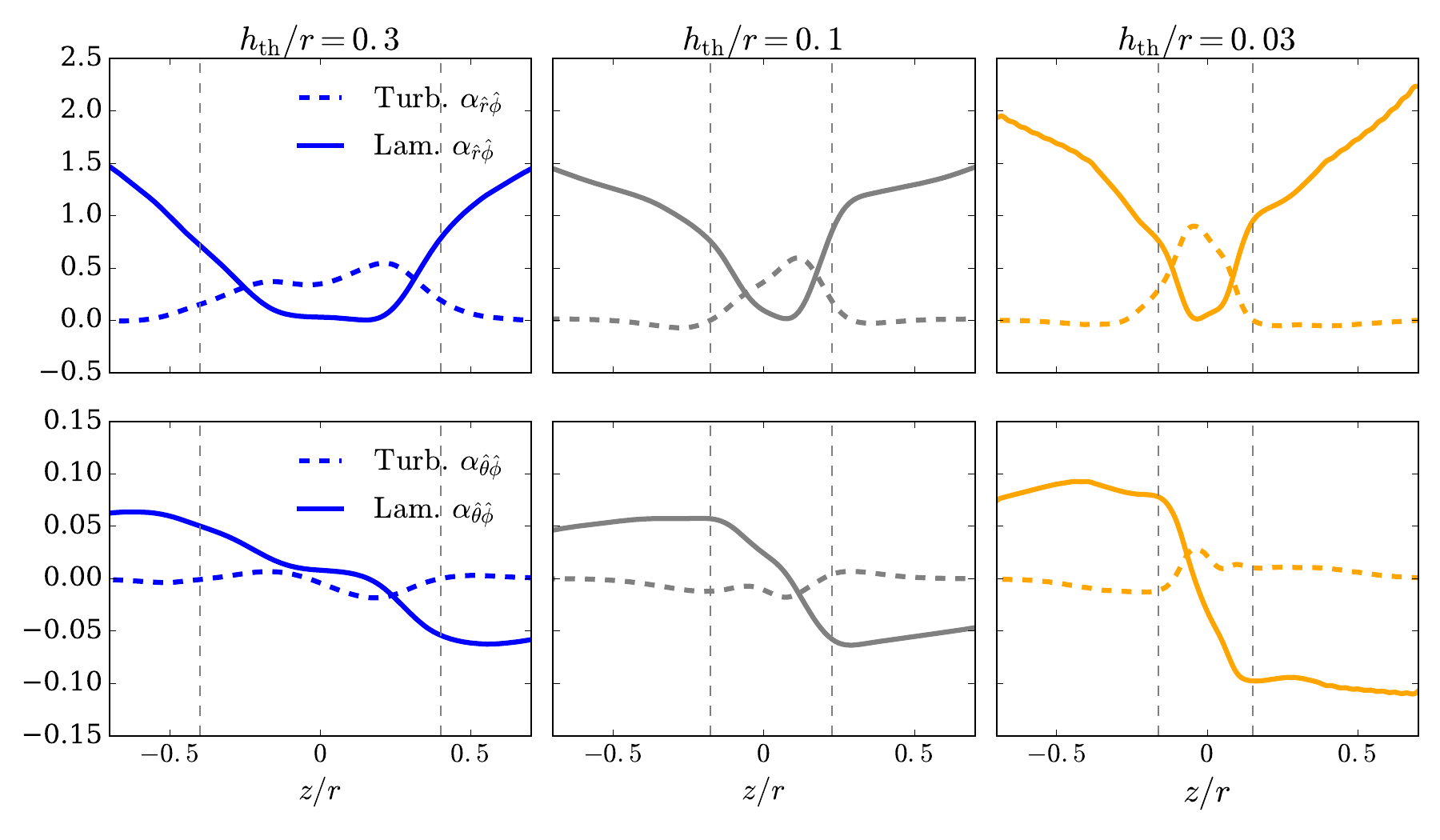}
\caption{Decomposition of the LNRF $r\phi$ Maxwell stress (top panels) and the LNRF $\theta\phi$ Maxwell stress (bottom panels) into a turbulent stress (dashed) and a laminar stress (solid) plotted as a function of $z/r$. The left, middle and right panels show the results for $h_\mathrm{th}/r=0.3, 0.1$ and $0.03$, respectively. All stresses are normalized by the midplane gas pressure. The grey dashed vertical lines show the surface of the disc. The plots are made at $r=7\:r_g$.}
\label{fig:stresses}
\end{figure*}

In  \autoref{fig:pressure_dec}, we see that for $h_\mathrm{th}/r=0.3$, the density profile is roughly in agreement with the expected Gaussian profile although the width of the density profile is narrower than expected. This is due to the jet compressing the disc as we will see on Figure \ref{fig:force_theta}. On the contrary, in the case of $h_\mathrm{th}/r=0.03$, and also to some extent for $h_\mathrm{th}/r=0.1$, the density is much broader than the expected Gaussian profile and follows a power-law going as $\sim z^{-4}$ at large $z$. We also see that the midplane density is much lower (by as much as two orders of magnitude for $h_\mathrm{th}/r=0.03$) than in the equivalent standard disc. 

To understand why the density scale height is so large in our $h_\mathrm{th}/r=0.03$ simulation, we plot in \autoref{fig:force_theta} the latitudinal force budget for all of our simulations. To do so, we rewrite the $\theta$-component of \autoref{eq:stress_energy_tensor}, for the latitudinal equilibrium of the disc, in the form
\begin{align}\label{eq:lat_eq}
    &-\partial_\mu(\sqrt{-g}(\rho h +2p_\mathrm{mag})u^\mu u_\theta)+\partial_\mu(\sqrt{-g}b^\mu b_\theta) \\ &\qquad +\sqrt{-g}\Bigl[-\partial_\theta(p_g+p_\mathrm{mag}) + \Gamma^\mu_{\mu \theta}(\rho h +2p_\mathrm{mag})u^\mu u_\mu \Bigr. \nonumber \\ 
    & \Bigl. \qquad\qquad - \Gamma^\mu_{\mu \theta}
    b^\mu b_\mu + \Gamma^t_{\theta\theta}T^\theta_t + \Gamma^r_{\theta\theta}T^\theta_r + \Gamma^\theta_{r\theta}T^r_\theta + \mathcal{O}(\frac{a}{M})\Bigr] =0.\nonumber
\end{align}
We associate the first term with the poloidal acceleration of the flow, the second term with the poloidal magnetic tension of the field lines, the third term with the gradient of the thermal and magnetic pressure, the spatial component of the fourth term with centrifugal-like effects, the temporal component of the fourth term with gravity-like effects, and the fifth term with effects of magnetic hoop stress. The remaining terms (dependent on the stress-energy tensor) vanish in the non-relativistic limit far from the black hole, and do not have Newtonian equivalents that we can easily interpret. We plot on \autoref{fig:force_theta} the sum of all the terms in \autoref{eq:lat_eq} as a black solid line. We do not plot separately the first, second, fifth, sixth, seventh, eight and ninth terms since they play a minor role in the latitudinal equilibrium, but we do take all of these terms into account in the sum.  In a steady state, the sum of all the terms in \autoref{eq:lat_eq} should vanish inside the disc. We see that this is approximately the case in all of our simulations.

From \autoref{fig:force_theta}, we see that for $h_\mathrm{th}/r=0.3$ the thermal pressure is the main force opposing the combined effects of the centrifugal acceleration and the gradient of laminar magnetic pressure. The laminar magnetic pressure force, supplied by the jet and the magnetic wind launched from the disc, is directed toward the midplane and compresses the disc. At $h_\mathrm{th}/r=0.1$, the gradient of thermal pressure is of the same amplitude as the gradient of turbulent magnetic pressure. At $h_\mathrm{th}/r=0.03$, the gradient of turbulent magnetic pressure is now the main force supporting the vertical structure and the gradient of thermal pressure plays a subdominant role. In \autoref{fig:pressure_dec}, we decompose the toroidal magnetic field pressure, which is the largest component of the field, into a laminar and a turbulent contribution. We see that the laminar magnetic field always approaches zero at the midplane because of an equatorial current sheet. As a result, the laminar magnetic field always compresses the disc regardless of whether it is attached to the jet or the disc. However, near the midplane the turbulent magnetic field always dominates over the laminar magnetic field and decreases with increasing $z$, providing latitudinal support to the disc body in addition to the thermal pressure. For $h_\mathrm{th}/r=0.03$, the turbulent magnetic pressure in the midplane is larger than the thermal pressure and so dominates the support of the thin disc structure, making it thicker than the thermal scale height (see also \autoref{fig:force_theta}). 

Finally, we note that we do not observe very strong elevated  accretion features as reported in the literature \citep{zhu2018,mishra2020,jacquemin2021}, where most of the accretion occurs at large height with $z\sim R$. This is consistent with \cite{jacquemin2021}, who showed that the height of the elevated accretion features decreased with magnetization before merging with the body of the disc. Our simulations have extreme magnetizations with $\beta_\theta$ as low as unity for $h_\mathrm{th}/r=0.03$, much lower than in \cite{mishra2020} where $\beta_\theta$ is around 100 for $h_\mathrm{th}/r=0.05$. Hence, our results confirm the trend that elevated accretion disappears at very high magnetizations. We note that for each simulation we did observe elevated accretion layers at earlier times, when the magnetization was lower. These features disappeared as time passed because of the disc becoming MAD and so increasing in magnetization. That said, the accretion in thin MADs always appears in a more extended region of $\theta$ around the midplane that in standard theory. For example, for $h_\mathrm{th}/r=0.3$ and $0.1$ we find that $63\%$ (for $1-e^{-1}$) of the accretion is happening inside $z/r\approx0.24$ and $0.11$ respectively; most of the accretion happens inside $h_\mathrm{th}/r$ in thick MADs. However, for $h_\mathrm{th}/r=0.03$, we find that $63\%$ of the accretion happens inside $z/r\approx0.10$.  In this sense, we can talk about elevated accretion for thin MADs, although we emphasize that it is not accretion happening primarily at very large height of $z\sim R$, as initially reported by \cite{zhu2018}, but rather over a wide range of latitude around the midplane.

\subsection{Angular momentum transport}\label{sec:angular_momentum}

\begin{figure}
\includegraphics[trim={0 10mm 0 0},width=80mm]{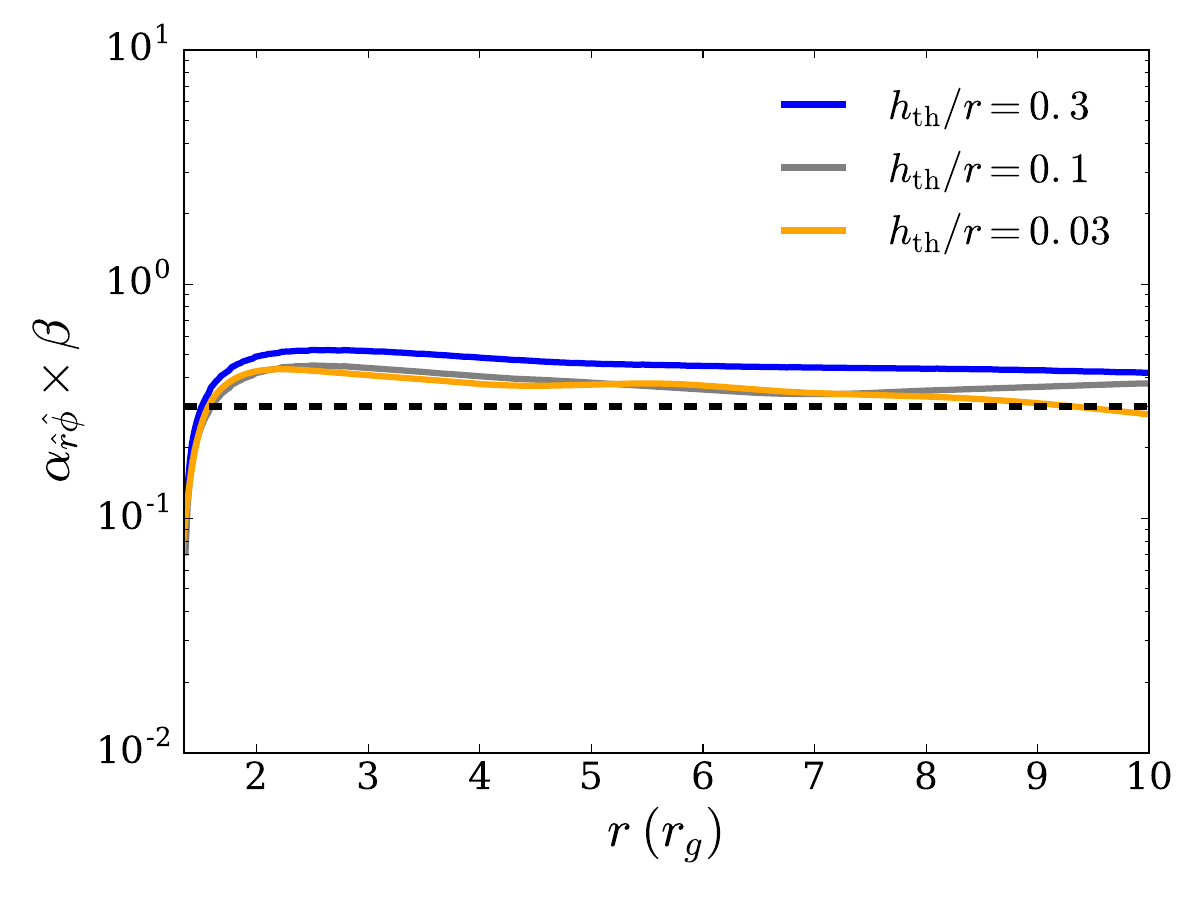}
\caption{$\alpha_{\hat{r}\hat{\phi}}\times\beta$ as a function of radius for all our MAD simulations. The dashed black line shows the expected value for MRI turbulence. The agreement between the expected value and our measured values of $\alpha_{\hat{r}\hat{\phi}}\times\beta$ suggest that MRI is driving the turbulence in MADs.}
\label{fig:alpha}
\end{figure}

\begin{figure*}
\includegraphics[trim={0 10mm 0 0},width=1.\textwidth]{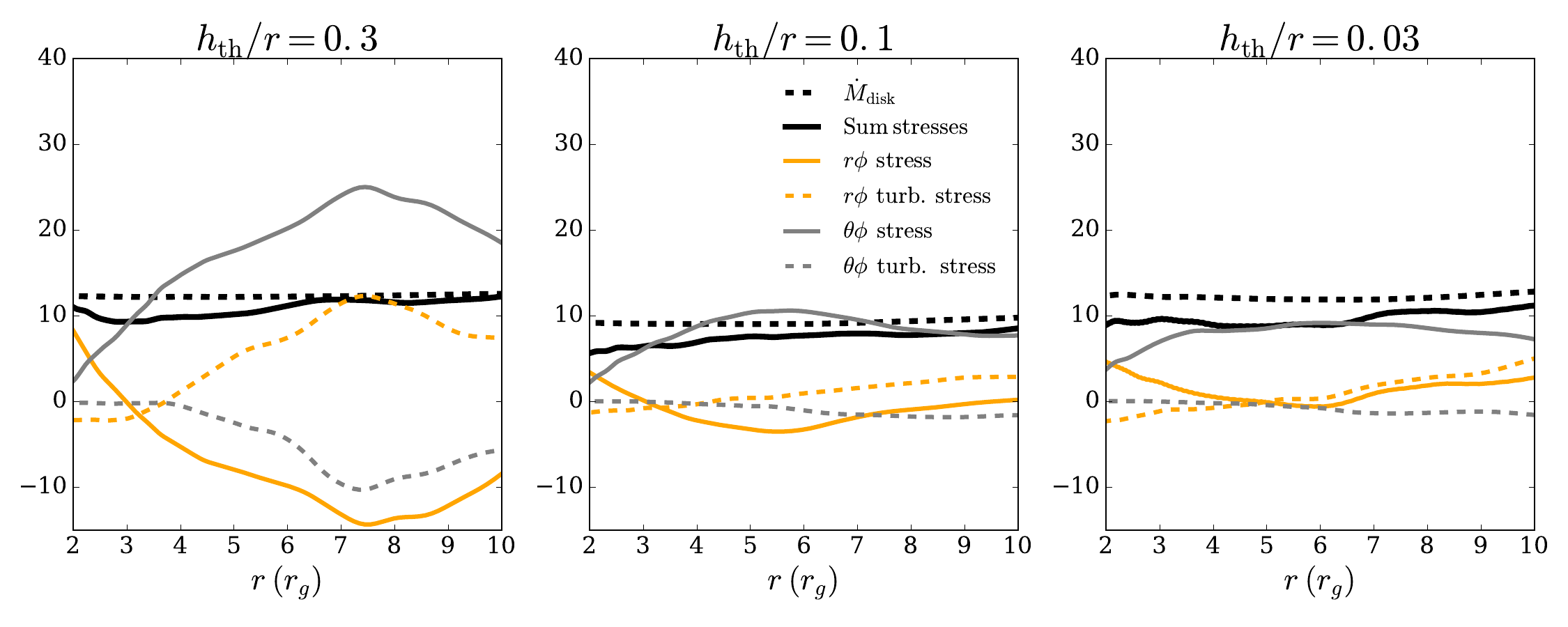}
\caption{Contributions of each component of the stress in \autoref{eq:mdot} to the total accretion rate as a function of radius. The orange lines show the contribution of the total $r\phi$ stress component (solid) and the turbulent $r\phi$ stress component (dashed). The grey lines show the contribution of the total $\theta \phi$ stress component (solid) and the turbulent $\theta\phi$ stress component (dashed). The solid black line shows the sum of all the components on the right hand side of \autoref{eq:mdot}. The dashed black line show our measurement of the mass accretion rate within the disc surface. The solid black line and the dashed black line should follow each other closely if we captured all the accretion mechanisms. The laminar $\theta\phi$ stress, the wind-driven stress, is always the main term driving accretion in MADs.}
\label{fig:ang_mom_transport}
\end{figure*}

\begin{figure}
\includegraphics[trim={0 10mm 0 0},width=90mm]{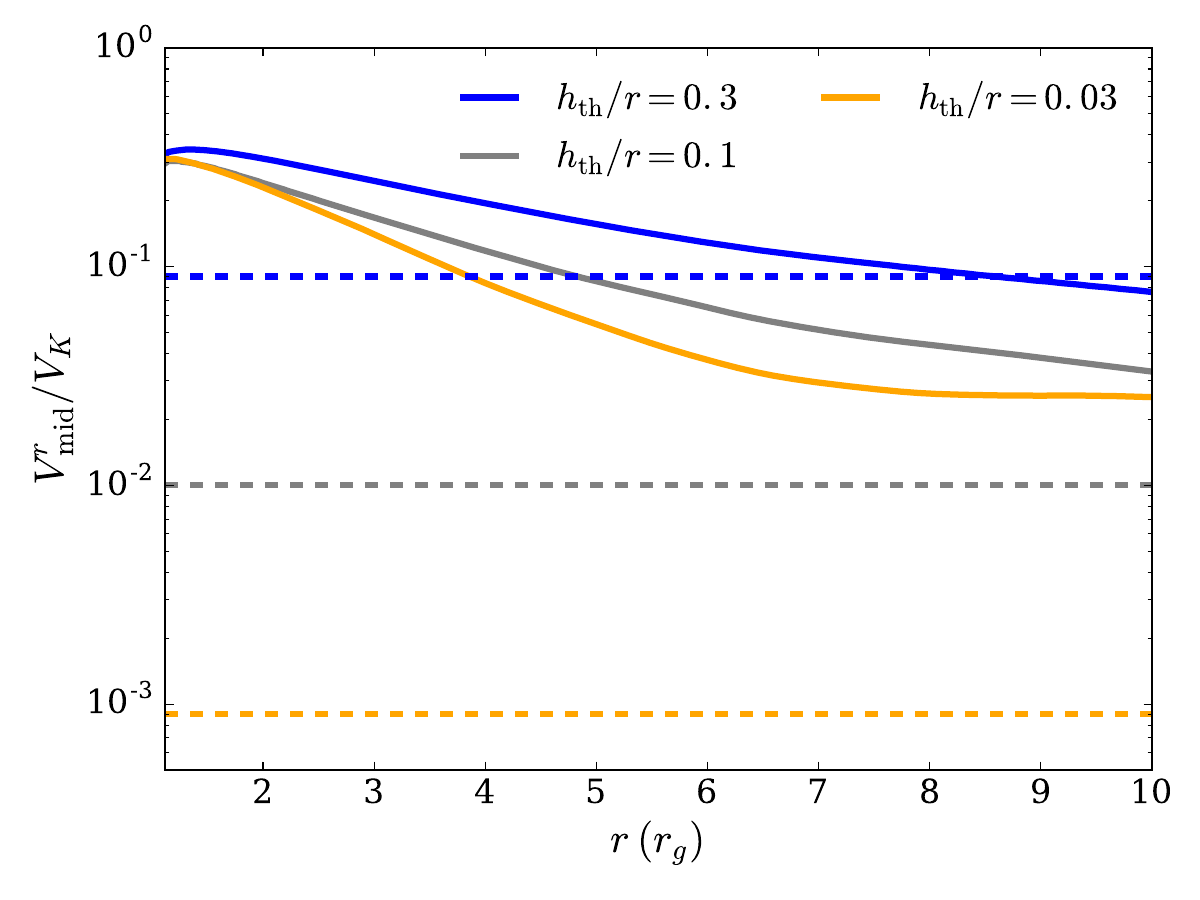}
\caption{Radial 3-velocity in the midplane normalized by the Keplerian velocity for different $h_\mathrm{th}/r$. For comparison, the dashed lines show the expectation from standard theory $V^r/v_\mathrm{K} = \alpha(h_\mathrm{th}/r)^2$, where we use $\alpha=1$. Our thin disc simulation has transonic/supersonic inflow speed, much higher than expected from standard theory.}
\label{fig:vr_r}
\end{figure}

We plot on \autoref{fig:stresses} the decomposition of the $r\phi$ Maxwell stress and the $\theta\phi$ Maxwell stress in the locally non-rotating frame (LNRF) as in \cite{novikov1973}, both normalized by the midplane thermal pressure, into turbulent and laminar components. The normalized turbulent $r\phi$ Maxwell stress corresponds to what is usually referred to as the $\alpha$-parameter \citep{Shakura}. Our turbulent values of $\alpha_{\hat{r}\hat{\phi}}$ are very high with values around $0.5$ for $h_\mathrm{th}/r=0.3$ and $0.1$ and as high as $\approx1$ for $h_\mathrm{th}/r=0.03$. These values are typical of what is found for MRI in the strong magnetization regime \citep{salvesen2016,scepi2018b}. Indeed, we show on \autoref{fig:alpha} the value of $\alpha_{\hat{r}\hat{\phi}}\times\beta$ as a function of radius for all of our simulations. We find that all MADs have $\alpha_{r\phi}\times\beta\approx 0.3-0.4$, very close to the expected value for MRI of $\alpha_{\hat{r}\hat{\phi}}\times\beta\approx0.3$ \citep{salvesen2016}. This suggests that MRI is driving the turbulence in all our MADs simulations. This is also consistent with what was found in a hot MAD with no cooling in \cite{begelman2022}. It is remarkable that the saturation level of MRI seems to depend mostly on $\beta$ and so little, if at all, on $h_\mathrm{th}/r$.

We also see from \autoref{fig:stresses} that the laminar $r\phi$ Maxwell stress is much lower than the turbulent one in the midplane but dominates at high latitudes. As expected from MRI, the $r\phi$ stress is positive meaning that angular momentum is transported outwards. We also see that the laminar $\theta\phi$ stress is negative above the midplane and positive below the midplane. This is consistent with angular momentum being transported from the disc to the wind. \autoref{fig:stresses} also shows that the $\theta\phi$ stress is always much smaller than the $r\phi$ stress. However it is not the stress by itself that induces accretion but the local divergence of the stress and we see from \autoref{fig:stresses} that the laminar $\theta\phi$ stress has a very large derivative in the $\theta$ direction.

We can compare the relative contributions of different stress components to the accretion by integrating the divergence of the stress over $\theta$. To do so, we combine \autoref{eq:continuity} and \autoref{eq:stress_energy_tensor} to write 
\begin{align}\label{eq:mdot}
\dot{M}_\mathrm{disc} &\equiv \int_{\theta_\mathrm{min}}^{\theta_\mathrm{max}}\int_\phi-\rho u^r \sqrt{-g}d\theta d\phi \nonumber \\
&= \int_{\theta_\mathrm{min}}^{\theta_\mathrm{max}}\int_\phi\frac{1}{u_{\phi,r}}\Big[\sqrt{-g}\rho u^\theta\partial_\theta(u_\phi)+\partial_r(\sqrt{-g} t^r_\phi) \nonumber \\
&\qquad  + \partial_\theta(\sqrt{-g} t^\theta_\phi) \Big]d\theta d\phi.
\end{align}
where we decomposed the stress-energy tensor into $T^{\mu\nu}=\rho u^\mu u^\nu + t^{\mu\nu}$, assumed axisymmetry, dropped the time derivative since we average over time and used the fact that $T^{\kappa}_{\lambda}\Gamma^{\lambda}_{\phi \kappa}=0$ because of the $\phi$-symmetry of the Kerr metric \citep{gammie2003}. The first term in the brackets on the right hand side of \autoref{eq:mdot} represents the mass lost in the wind transporting its own angular momentum. This term does not lead to accretion since it does not transport any extra angular momentum from the remaining material. It should be identically zero, close to what we find. The second term is the contribution of the radial torque decomposed into a turbulent and a laminar contribution. Note that the contribution of the radial torque is often reduced to the MRI-induced turbulent torque in standard theory \citep{Shakura} but large scale magnetic torques, that are associated with the large-scale structure of the wind, can also contribute to accretion \citep{blandford1982}. The third and last term is the contribution of the latitudinal torque, also decomposed into a turbulent and a laminar contribution.  We plot the total of all the terms as a black solid line on \autoref{fig:ang_mom_transport} but only show separately the most important contributions, i.e. the turbulent and laminar contributions of the $r\phi$ and $\theta\phi$ stresses. To obtain the total accretion rate in the disc, we need to integrate the divergence of the stress over the disc. As for \autoref{fig:pressure_dec} and \autoref{fig:stresses}, we define the surface of the disc as the place where the outflow crosses the slow magnetosonic point. We note that this surface follows closely the surface where the radial velocity is zero and so is a good proxy to locate the interface between the disc and the outflow. At small radii $\lesssim 3\:r_g$, where the radial velocity becomes negative all around the black hole, we define the surface of the disc by extrapolating inward the surface of constant $\theta$ that defines the disc surface at larger radii. 

 We see from \autoref{fig:ang_mom_transport} that for all $h_\mathrm{th}/r$, the accretion is largely dominated by the divergence of the laminar, large-scale stresses, and especially the contribution of the $\theta\phi$ stress to the divergence. The predominance of the large-scale stresses means that angular momentum transport in MADs is wind-driven and not MRI-driven. Again, although the amplitude of the $\theta\phi$ stress is small compared to that of the $r\phi$ stress, the contribution of the $\theta\phi$ stress to angular momentum transport is larger than that of the $r\phi$ stress. Note that it seems that the laminar $r\phi$ stress induces excretion in the simulations with $h_\mathrm{th}/r=0.3$ and $0.1$. This is an artifact of our decomposition of the divergence into $r\phi$ and $\theta\phi$ terms. What is physically relevant is the decomposition into a turbulent term, the MRI driving, and a laminar term, the wind driving. For all $h_\mathrm{th}/r$, the MRI-induced, turbulent stress contributes to  accretion but its contribution is always smaller than that from the laminar, wind-driven stress, which accounts for most of the accretion.
 
 We see from \autoref{fig:vr_r} that the accretion driven by the wind-driven torque is much faster than that predicted from standard theory, with supersonic accretion even for $h_\mathrm{th}/r=0.03$. This wind-driven accretion could explain why most, if not all, GRMHD simulations of MADs in the literature report effective $\alpha$ values (computed from the radial velocity) larger by almost an order of magnitude than the $\alpha_{\hat{r}\hat{\phi}}$ measured directly from the turbulence in those simulations \citep{avara2016,liska2022}. Finally, we reiterate that the much faster inflow driven by laminar fields is the main reason why the density in our simulations is much lower than expected from standard theory (see \S\ref{sec:vertical}).

\subsection{Dissipation }\label{sec:heating}

\subsubsection{Latitudinal profile}
\begin{figure}
\includegraphics[trim={0 10mm 0 0},width=90mm]{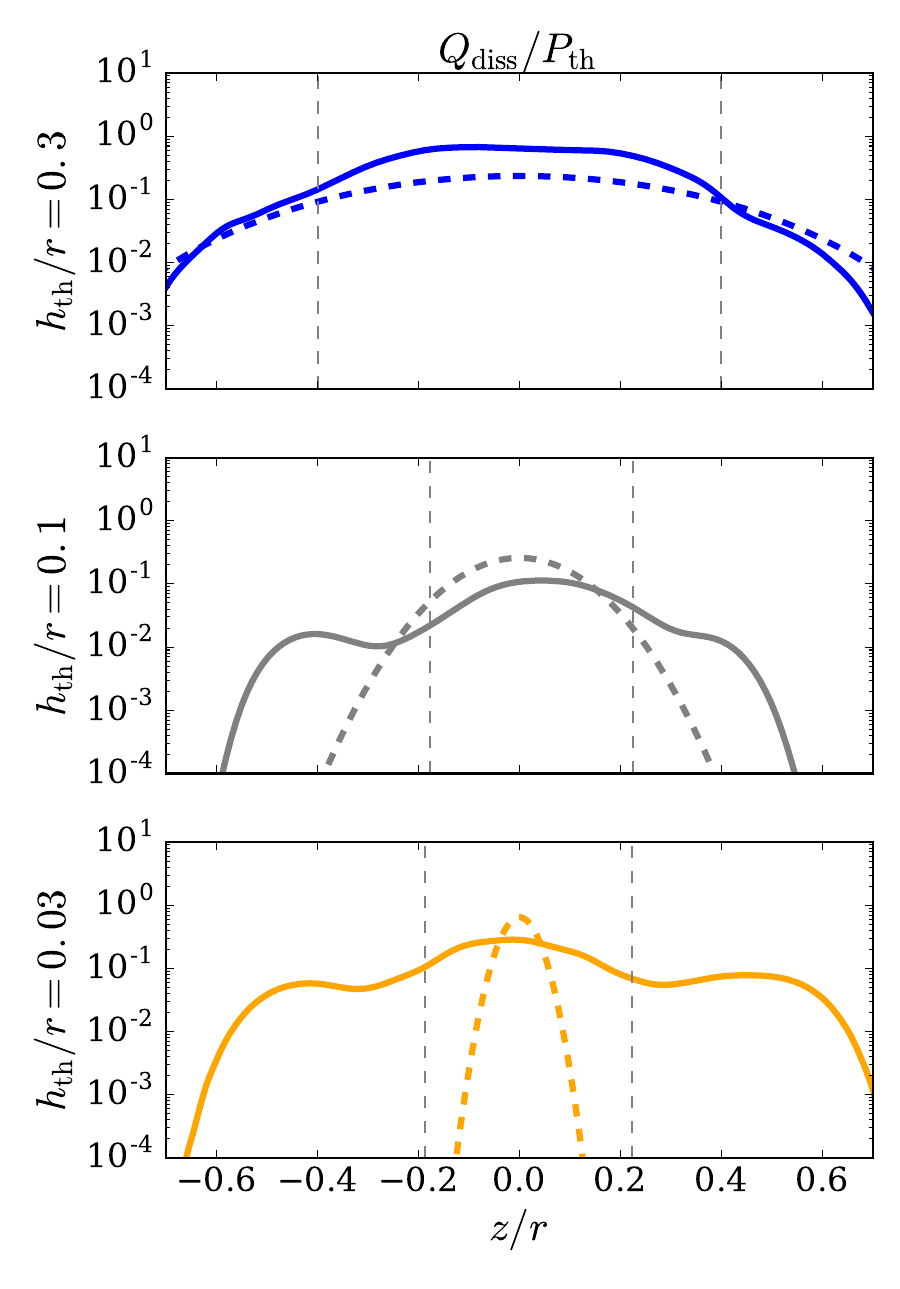}
\caption{The solid lines show the fluid-frame dissipation of heat measured in our simulation as a function of $z/r$ at $r=7\:r_g$. For comparison, the dashed lines show a Gaussian density profile of width $h_\mathrm{th}/r$, which is the expected dissipation profile of an isothermal disc supported by thermal pressure as in standard theory. The measured and expected dissipation profiles are normalized by the dissipation in the midplane of a standard isothermal disc accreting at the same accretion rate as in our simulation.  The grey dashed vertical lines show the surface of the disc. For the thin disc simulations the heating rate profile is much more extended in $\theta$ than expected from standard theory.}
\label{fig:heating_theta}
\end{figure}

The use of a cooling function allows us to measure the azimuthally averaged amount of energy per unit volume and proper time that is locally dissipated in the fluid-frame,
\begin{equation}
    Q_\mathrm{diss} \equiv \int_\phi \mathcal{F}_t \frac{u^t}{u_t}\alpha_\mathrm{lapse} d\phi,
\end{equation}
where $\alpha_\mathrm{lapse}$ is the lapse function. Indeed, the amount of energy that has to be removed to keep the gas at a specified temperature is the irreversible energy that is dissipated locally through shocks and turbulence plus the reversible work energy on the fluid minus the amount of energy that is globally transported. 

On \autoref{fig:heating_theta} we plot $Q_\mathrm{diss}$ as a function of $\theta$ for each of our MAD simulations. We also plot as dashed lines the dissipation profile that an isothermal, thermally supported disc with the same accretion rate would have in hydrostatic equilibrium. The expected dissipation profile is normalized to the standard midplane dissipation rate of a disc with the same accretion rate as in our simulation. We see that for $h_\mathrm{th}/r=0.3$, the dissipation profile follows closely the standard profile. This is because, in this case the disc is thermally supported in the latitudinal direction and its structure is close to standard theory. However, as we decrease $h_\mathrm{th}/r$ the disc becomes increasingly magnetically supported in the latitudinal direction (see \S\ref{sec:vertical}) and its structure deviates from standard theory. In the thin, magnetically supported cases, we see a lot of extra dissipation at higher latitudes, as can be seen on \autoref{fig:heating_theta} for $h_\mathrm{th}/r=0.1$ and $0.03$. This is especially striking for $h_\mathrm{th}/r=0.03$, where a large fraction of the dissipation is located outside the thermal density profile, creating a ``corona''. We note that this corona is partially composed of outflowing material from the disc resembling a wind (see \S\ref{sec:winds}).  For $h_\mathrm{th}/r=0.03$, we find that $\approx 31\%$ of the dissipation between $4$ and $10\:r_g$ occurs in the wind. We excluded regions below $4\:r_g$ because the jet is compressing the disc here and preventing the wind from being launched. The fraction of dissipation in the wind is $\approx5\%$ when taking into account all $r<10\:r_g$.

\begin{figure}
\includegraphics[trim={0 10mm 0 0},width=90mm]{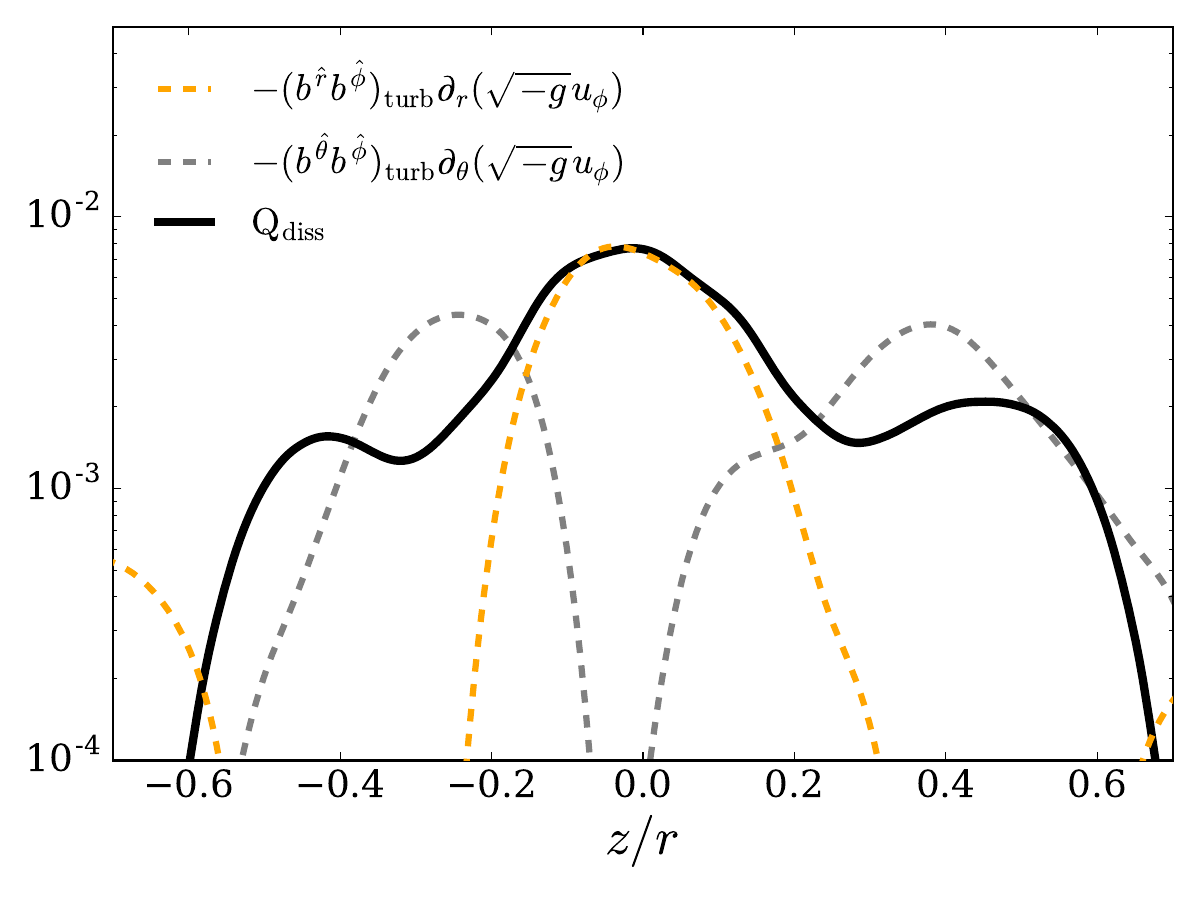}
\caption{Comparison of the fluid-frame dissipation measured by our cooling function (solid black line) to the turbulent dissipation associated with the turbulent $r\phi$ magnetic stress times the radial shear (dashed orange line) and the turbulent $\theta\phi$ stress times the latitudinal shear (dashed grey line) as a function of $z/r$ at $r=7\:r_g$ for the case $h_{\rm th} = 0.03$. The $r\phi$ and $\theta\phi$ turbulent stresses account within a factor of order unity for the midplane and coronal dissipation,  respectively.}
\label{fig:heating_dec}
\end{figure}

To understand where this extra coronal emission is coming from in our thinnest case we plot the contribution of the stresses to the dissipation of energy as in Eq. 17 of \cite{beckwith2008}, where we added the influence of the $\theta\phi$ stress to write\footnote{Note that there should be a term in \autoref{eq:diss_turb} taking into account the dissipation of the laminar magnetic field energy due to turbulent resistivity. Computing this term is beyond the scope of this paper given the complexity of measuring the electromagnetic force to a high accuracy. We remark however that our estimate of the dissipation in \autoref{eq:diss_turb} seems to already be able to account for the measured dissipation rate of energy suggesting that the resistive heating might be subdominant.}
\begin{equation}\label{eq:diss_turb}
    Q_\mathrm{diss} \approx -(b^{\hat{\phi}}  b^{\hat{r}})_\mathrm{turb} \partial_r(u_\phi) - (b^{\hat{\phi}} b^{\hat{\theta}})_\mathrm{turb} \partial_\theta(u_\phi).
\end{equation}
 We see from \autoref{fig:heating_dec} that the energy deposited locally comes from two sources: the turbulent $r\phi$ stress extracting energy from the radial angular velocity gradient and the turbulent $\theta\phi$ stress extracting energy from the latitudinal angular velocity  gradient. However, the latitudinal profiles of the two contributions are very different. The contribution from the turbulent $r\phi$ stress is concentrated around the midplane with $z/r \lesssim 0.2$ while the contribution from the turbulent $\theta\phi$ stress peaks between $z/r\approx 0.3-0.5$. This difference in latitudinal profiles can explain the latitudinal profile of the measured dissipation in our simulation, which is formed of one bump in the midplane, the contribution from the $r\phi$ stress, and two off-centered bumps, the contribution from the $\theta\phi$ stress. We note that the dissipated energy estimated by measuring the stresses is higher by a factor $\approx2$ than the measured dissipated energy. This might be due to the fact that measuring dissipation through the cooling function does not differentiate between irreversible and reversible energy exchange (here adiabatic cooling in the expanding outflow).

\begin{figure}
\includegraphics[trim={0 10mm 0 0},width=85mm]{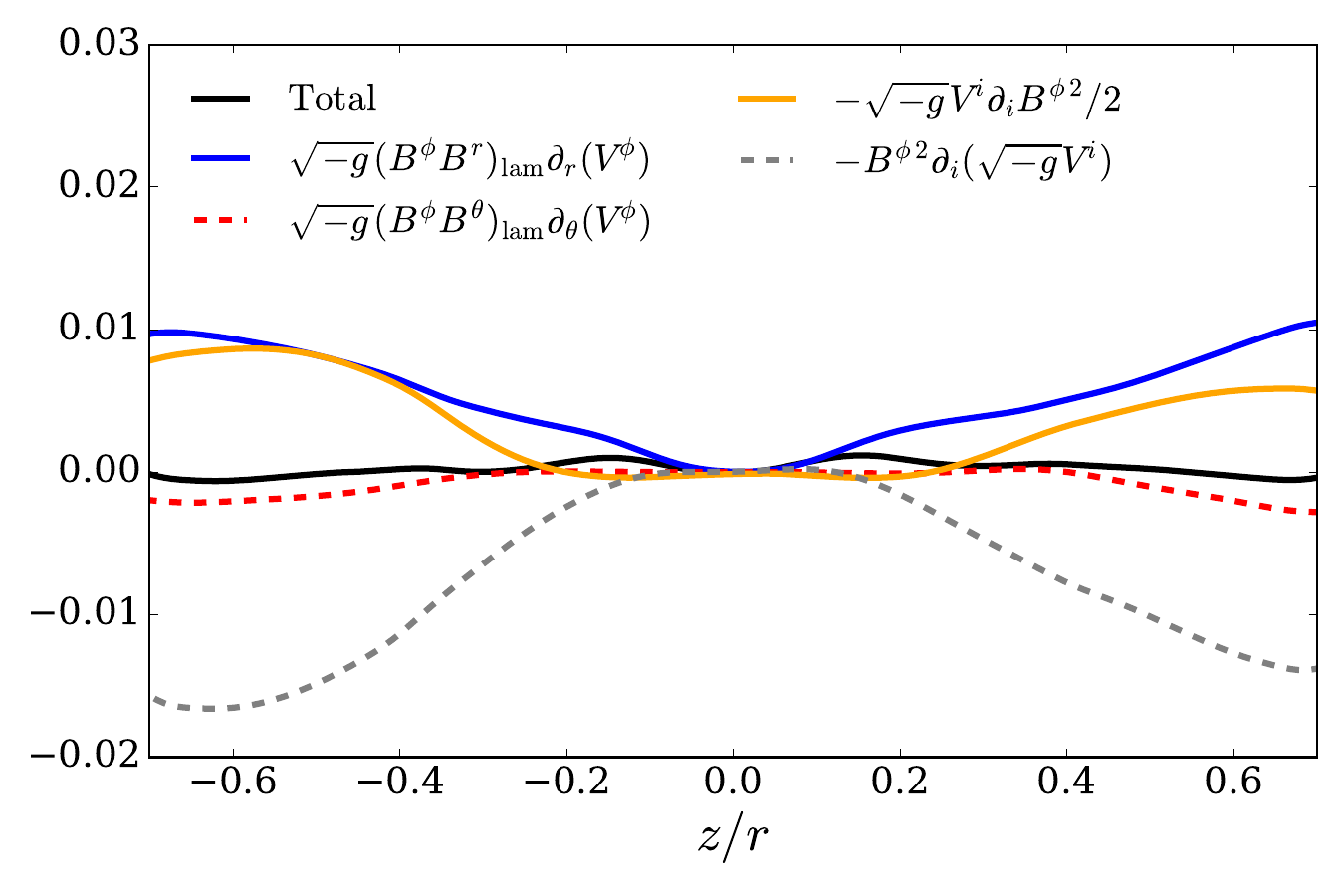}
\caption{Individual terms in \autoref{eq:diss_lam} as a function of $z/r$ at $r=7\:r_g$ for our simulation with $h_\mathrm{th}/r=0.03$. The solid blue, dashed red, solid orange and dashed grey lines show the contribution of the laminar $r\phi$ stress, the laminar $\theta\phi$ stress, the covariant derivative of the laminar toroidal magnetic energy and the work by the laminar toroidal magnetic energy respectively. The solid black line shows the sum of all these contributions. The fact that the sum is zero shows that the laminar stresses do not dissipate heat in the disc, in contrast to the turbulent stresses shown in \autoref{fig:heating_dec}.}
\label{fig:diss_lam}
\end{figure}

Finally, we note that although angular momentum is mainly transported by the laminar Maxwell stresses (especially the laminar Maxwell $\theta\phi$ stress), the dissipation of energy is only related to the sub-dominant turbulent components of the stresses. This is because the energy liberated by the laminar stresses mainly goes into Poynting flux and kinetic energy in the wind (neglecting the resistive dissipation of the field) \citep{blandford1982,ferreira1995}. This can be seen by taking the $\phi$ component of the induction equation (\autoref{eq:induction}) in the steady-state approximation and rewriting it as a function of the 3-vector magnetic field, $B^i$, and the 3-vector velocity, $V^i$, to find
\begin{equation}\label{eq:diss_lam}
    \sqrt{-g} (B^{\phi}B^{i})_\mathrm{lam}\partial_i V^\phi - (B^{\phi})^2\partial_{i}(\sqrt{-g}V^{i})-\sqrt{-g}V^{i}\partial_i (\frac{(B^{\phi})^2}{2}) = 0,
\end{equation}
where we used 3-vectors instead of 4-vectors to get as close as possible to a classical physics expression. The first term is the laminar equivalent of the turbulent dissipative terms in \autoref{eq:diss_turb}, which extracts energy from the angular velocity gradient. \autoref{eq:diss_lam} tells us that in ideal MHD the energy extracted from the laminar stresses and stored inside the $B_\phi$ laminar field should be compensated by work done by the toroidal magnetic field on accelerating the outflow (second term) and by magnetic energy advection in the outflow, i.e. goes into Poynting flux. We see from \autoref{fig:diss_lam} that \autoref{eq:diss_lam} is relatively well-verified in our simulation, illustrating that virtually none of the energy that is extracted by the laminar stresses is locally dissipated as heat. Contrary to the laminar terms, the turbulent terms act as an effective viscosity and resistivity in the fluid and so are the terms responsible for irreversible dissipation \citep{balbus1999}.

\subsubsection{Radial profile}
\begin{figure}
\includegraphics[trim={0 10mm 0 0},width=90mm]{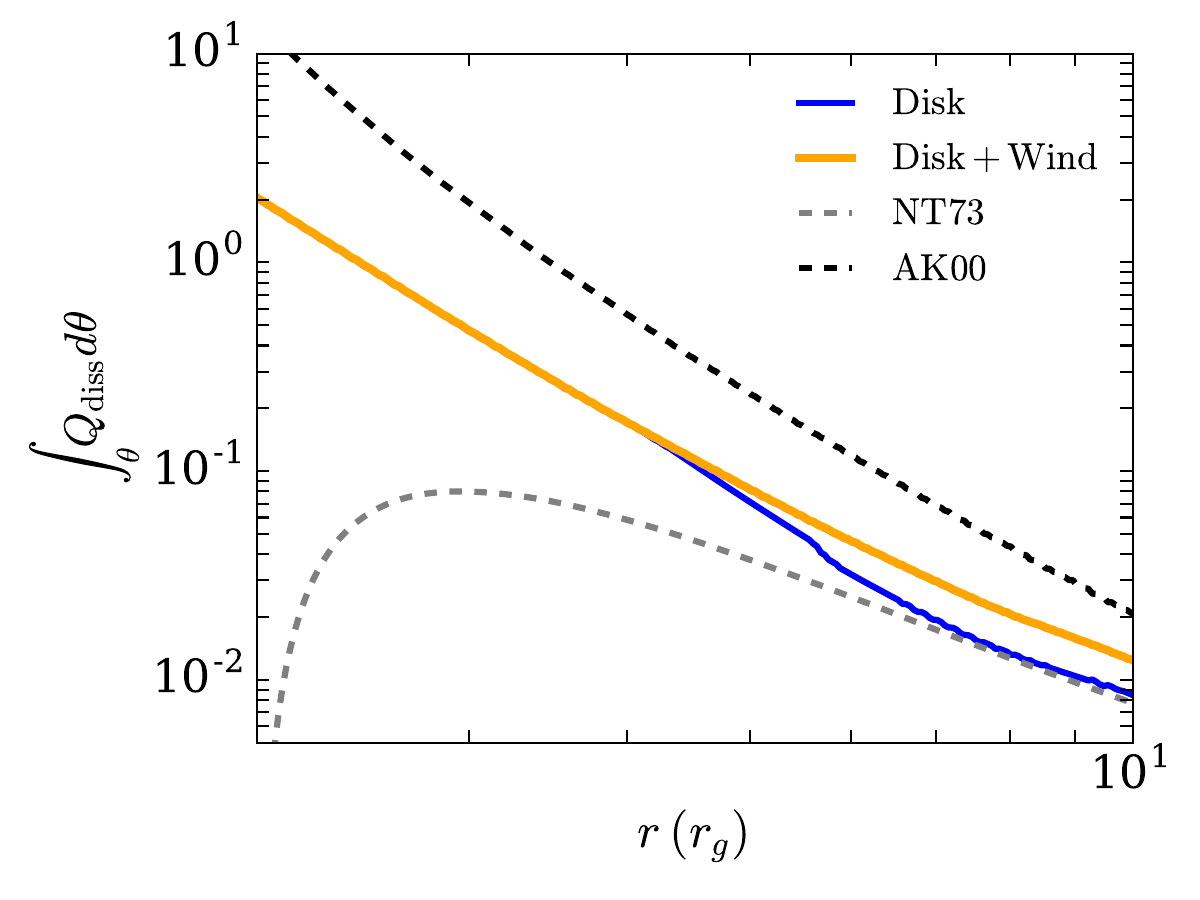}
\caption{Fluid-frame dissipation rate integrated along $\theta$ and $\phi$ as a function of $r$ for our simulation with $h_\mathrm{th}/r=0.03$. The yellow line shows the total dissipation in the disc+wind while the blue line shows the dissipation in the disc. The grey and black dashed line respectively show the expected profile from the standard models of NT73 and AK00. Our simulations show a lot of dissipation in the plunging region, because of the presence of a strong inner magnetic torque, so that it dissipates more than expected from NT73. However, because of the wind-driven accretion, the disc is less dissipative than it should be when comparing to AK00, which takes into account the effect of the inner torque.}
\label{fig:heating_r}
\end{figure}

We focus on this subsection on the radial profile of energy dissipation in our thinnest simulation, since it is the one that can be compared with standard thin disc models. We plot on \autoref{fig:heating_r} the latitudinally integrated, fluid-frame dissipation rate as a function of radius, for the disc alone (blue solid line)\footnote{Note that the boundary between the disc, which is defined here as the inflowing region with a magnetization smaller than unity, and the jet/wind is somewhat artificial near the BH. This is why we observe a sharp drop in the disc dissipation rate near $4\:r_g$.}, and also for the combined  disc and wind regions, i.e., including the outflowing regions (orange solid line). For comparison, we also plot the expected profile of dissipation from the \cite{novikov1973} model (NT73; grey dashed line) and the \cite{agol2000} model (AK00; black dashed line), which is an extension of the \cite{novikov1973} model taking into account a non-zero stress at the innermost stable circular orbit (ISCO) \citep{gammie1999,krolik1999}. The inner stress is a free parameter in AK00, which we fix by computing the inner stress from our simulation as $\int_\phi\int_{\theta_\mathrm{min}}^{\theta_\mathrm{max}} -(b^r b_\phi)_\mathrm{ISCO}\sqrt{g_{\theta\theta}}d\theta d\phi$ to match the definition in AK00.

We see from \autoref{fig:heating_r} that our simulation exhibits a lot of dissipation inside the ISCO, as was already noted in a number of magnetized simulations \citep{krolik2005,beckwith2008,noble2009,penna2010,avara2016}. Because of this additional inner stress, the disc is more radiatively efficient than would be expected from the NT73 model. However, we also note that our thin disc dissipates less energy than the AK00 model would predict using the inner stress computed from our simulation. This is expected from a disc where angular momentum transport is mostly driven by a wind-driven stress. Indeed, if the disc were entirely wind-driven, the gravitational energy would ultimately flow into the wind and not be dissipated locally\footnote{Note that a turbulent resistivity is always necessary to maintain the magnetic structure and allow the magnetic field lines to partially slip through the accreted matter. This turbulent resistivity will be accompanied by local dissipation of energy in the disc so that a wind-driven disc can never be purely dissipationless \citep{ferreira1995}.} \citep{blandford1982,ferreira1995} as opposed to a purely turbulent disc where the gravitational energy is transported outwards and then deposited locally at a rate of three times the local binding energy \citep{novikov1973}. We found in \S\ref{sec:angular_momentum} that, in our thin MAD simulation, angular momentum transport due to the turbulent radial stress was able to account for at most a third of the accretion while the rest was driven by large-scale stresses. This is consistent with the dissipation in the disc being at most a third of the expected dissipation from AK00. Now, some dissipation (roughly $30\%$ of the total dissipation) also happens in the wind, i.e. a part of the energy extracted by the wind is eventually dissipated as heat too. We find that this fraction is also consistent with the disc+wind emission being at most half of the expected dissipation from AK00.

To compute the radiative efficiency of the disc we time- and $\phi$-average our dissipation profile and integrate over $\theta$  to obtain a 1D effective dissipation rate as a function of radius as in \autoref{fig:heating_r}. We then use the ray-tracing code {\fontfamily{qcr}\selectfont grtrans} \citep{dexter2009,dexter2016} to estimate the radiative efficiency for an observer at infinity. To do so, we modified the emissivity of the disc in the disc test problem of {\fontfamily{qcr}\selectfont grtrans} but otherwise kept the observer configuration presented in \cite{dexter2009}. Note that with our procedure we do not take properly into account the relativistic effects for the emission originating in the wind since we flattened the emission profile in a razor-thin disc geometry. We believe that this is enough for our comparison since many other factors, such as the limited radial extent of the disc or the realistic capacity of the wind to radiate its energy, impede a precise estimate of the radiative efficiency. After ray-tracing, we find a radiative efficiency of $\approx32\%$ compared to $\approx18\%$ for a Novikov-Thorne disc. This is an enhancement of $\approx78\%$ compared to standard theory. Interestingly, this number is very close to \cite{avara2016}, who find an enhancement of $80\%$ but for a spin of 0.5 compared to 0.9375 in our case. We note that, because of the heavy use of floors and ceilings stabilizing the numerical scheme in the jet, we are also not taking into account dissipation of energy inside the jet which could also lead to extra dissipation compared to standard models. 

\subsection{Jet and wind efficiencies}\label{sec:winds}

\begin{figure}
\includegraphics[trim={0 10mm 0 0},width=90mm]{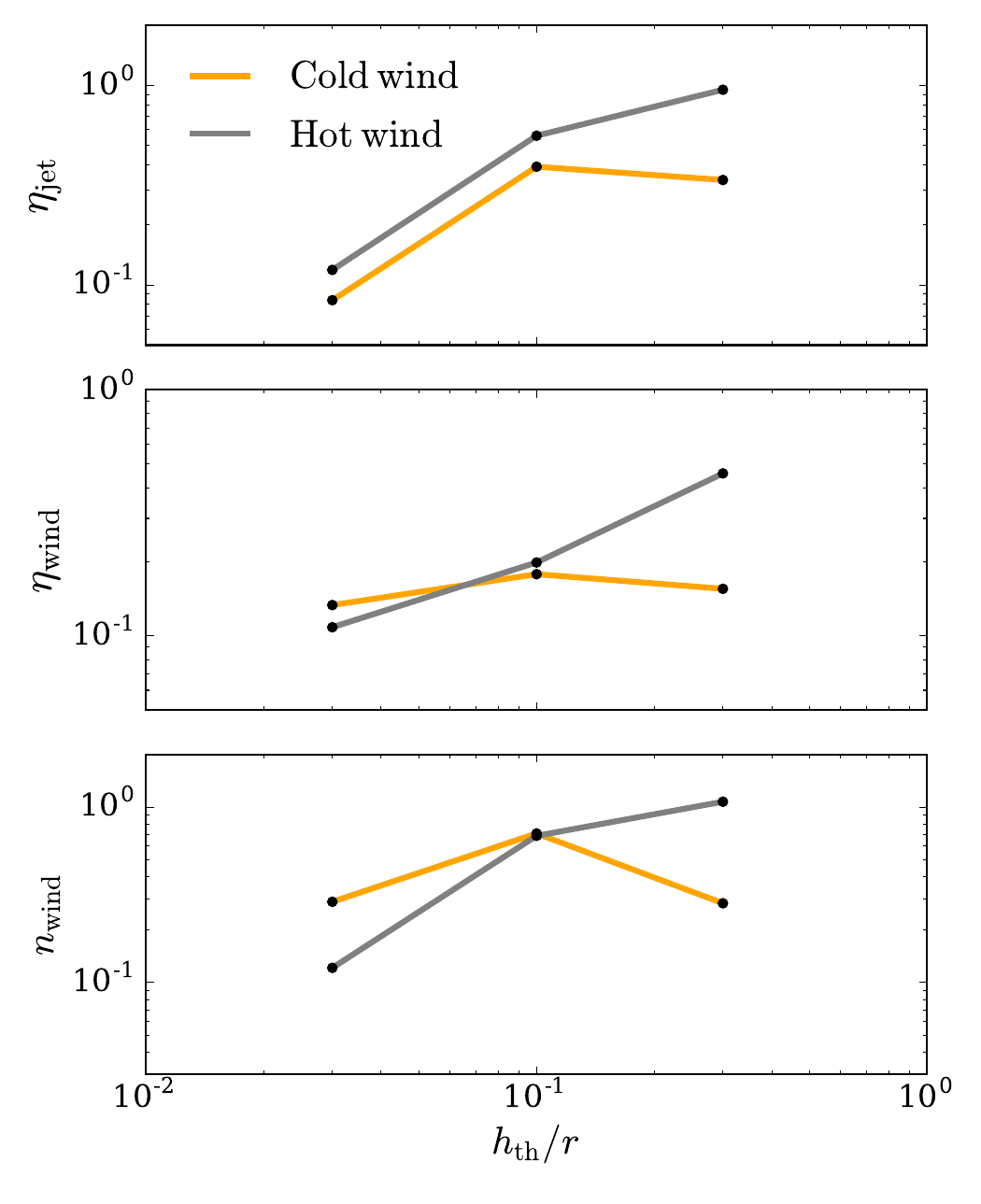}
\caption{Jet energy efficiencies (top panel), wind energy efficiencies (middle panel) and wind mass loss rate efficiencies (bottom panel) as a function of $h_\mathrm{th}/r$ at $r=40\:r_g$. For each panel, we show the results when the wind is cooled (orange line) or not (grey line). The efficiencies for $h_\mathrm{th}/r=0.3$ depend a lot on whether the wind is cooled or not.}
\label{fig:efficiencies}
\end{figure}

We show on \autoref{fig:efficiencies} the energy extraction efficiency of the jet (top panel) and the wind (middle panel) for our three simulations. The efficiencies are computed as 
\begin{equation}
    \eta_\mathrm{jet/wind} = \frac{\int_\mathcal{S_\mathrm{jet/wind}} \sqrt{-g}(T^r_t+\rho u^r)d\theta d\phi}{\dot{M}}
\end{equation}
where the surface of integration $\mathcal{S}_\mathrm{jet/wind}$ is defined as the surface on a sphere of radius $r=40\:r_g$ where $\sigma>1$ for the jet and where $\sigma<1$ and $u^r>0$ for the wind. As explained in \S\ref{sec:methods}, we show on  \autoref{fig:efficiencies} the results of two sets of simulations for each $h_\mathrm{th}/r$: one set where the wind is cooled, shown as an orange line, and one set where the wind is not cooled, shown as a grey line. These two sets allow us to study, in a very simplistic way, the effect that the gas temperature of the wind has on the outflow properties.

We see from the top panel of \autoref{fig:efficiencies} that when the wind is hot the efficiency of the jet is proportional to the first power of $h_\mathrm{th}/r$. It goes from $\approx 10\%$ for $h_\mathrm{th}/r=0.03$ to $\approx 100\%$ for $h_\mathrm{th}/r=0.3$. A jet efficiency larger than $100\%$ is usual in thick discs, where it is interpreted as the jet tapping into the rotational energy of the black hole \citep{tchekhovskoy2011}. Our value of $10\%$ for the thin disc is also in agreement with the literature, with reported values of $20\%$ for the same spin and $h_\mathrm{th}/r$ in \cite{liska2022}.  The increase with $h_\mathrm{th}/r$ is consistent with the fact that the magnetic flux on the black hole, $\Phi_\mathrm{sph}$, increases as $(h_\mathrm{th}/r)^{1/2}$ as can be seen on the first panel of  \autoref{fig:phi_beta}. We note that our scaling of $\eta_\mathrm{jet}\propto h_\mathrm{th}/r$ is different from the scaling $\eta_\mathrm{jet}\propto (h_\mathrm{th}/r)^2$ found in \citep{avara2016}. It is unclear where this difference comes from. 

We see in the top panel of  \autoref{fig:efficiencies} that the jet efficiency is always smaller when the wind is cold. This is especially striking for the hot MAD with $h_\mathrm{th/r}=0.3$ where the efficiency goes from $\approx100\%$ to $\approx30\%$. This comes from the fact that the flux on the black hole decreases when the wind is cooled, thus decreasing the jet power as we showed in \S\ref{sec:time_dependent}.  This effect is much less pronounced when the disc becomes thinner, with an increased efficiency of only $30\%$ between the hot wind case and cold wind case for $h_\mathrm{th/r}=0.1$ and 0.03.

Interestingly, we see from the middle panel of \autoref{fig:efficiencies} that the wind efficiencies do not depend as strongly on $h_\mathrm{th}/r$ as the jet efficiencies. This is especially true in the case where the wind is cold, where the efficiency is almost constant with $\eta_\mathrm{wind}\approx 15-20\%$. The only case where the wind energy efficiency varies significantly is when the disc is hot, with $h_\mathrm{th}/r=0.3$, and the wind is hot too. In this case, the efficiency increases up to $\approx 50\%$.  It is worth noting that for $h_\mathrm{th}/r=0.03$ the efficiency of the wind becomes larger than the efficiency of the jet. These wind efficiencies are in good agreement with those found in \cite{liska2022}. 

We show on the bottom panel of \autoref{fig:efficiencies} the wind mass loss rate of the disc per unit of logarithmic radius defined through the parameter $n_\mathrm{wind}$ as in \cite{blandford2004},
\begin{equation}
    n_\mathrm{wind} = \frac{d \ln \dot{M}}{d\ln r} =  \frac{[\int_\phi \rho u^r \sqrt{-g} d\theta d\phi]^{\theta_\mathrm{max}(r)}_{\theta_\mathrm{min}(r)}}{\dot{M}}.
\end{equation}
Note that $0<n_\mathrm{wind}<1$ if there is no other source of energy than the accretion. The wind mass loss rate is measured at the surface of the disc defined by $\theta_\mathrm{max}(r)$ and $\theta_\mathrm{min}(r)$, which are the angles where the flow reaches the slow magnetosonic point for each radius. We also looked at the wind mass loss rate due to $\rho u^\theta$ but find that the latitudinal term is always smaller than the radial term and so we do not show it here. 

The first thing we find is that $n_\mathrm{wind}$ is roughly constant as a function of radius, although we do not show it here, suggesting an almost self-similar disc. We then see from \autoref{fig:efficiencies} that, when the wind is not cooled, the local wind mass loss rate goes from weak with $n_\mathrm{wind}\approx 0.12$ for $h_\mathrm{th}/r=0.03$ to very strong with $n_\mathrm{wind}\approx 1.07$ for $h_\mathrm{th}/r=0.3$. This is not the case when the wind is cold, where we observe a maximum for $h_\mathrm{th}/r=0.1$ similar to what is observed for the wind energy efficiencies. Understanding why and how the wind efficiency and mass loss rate vary with the thermal content of the wind and its magnetization is beyond the scope of this paper.

\begin{figure}
\includegraphics[trim={0 10mm 0 0},width=90mm]{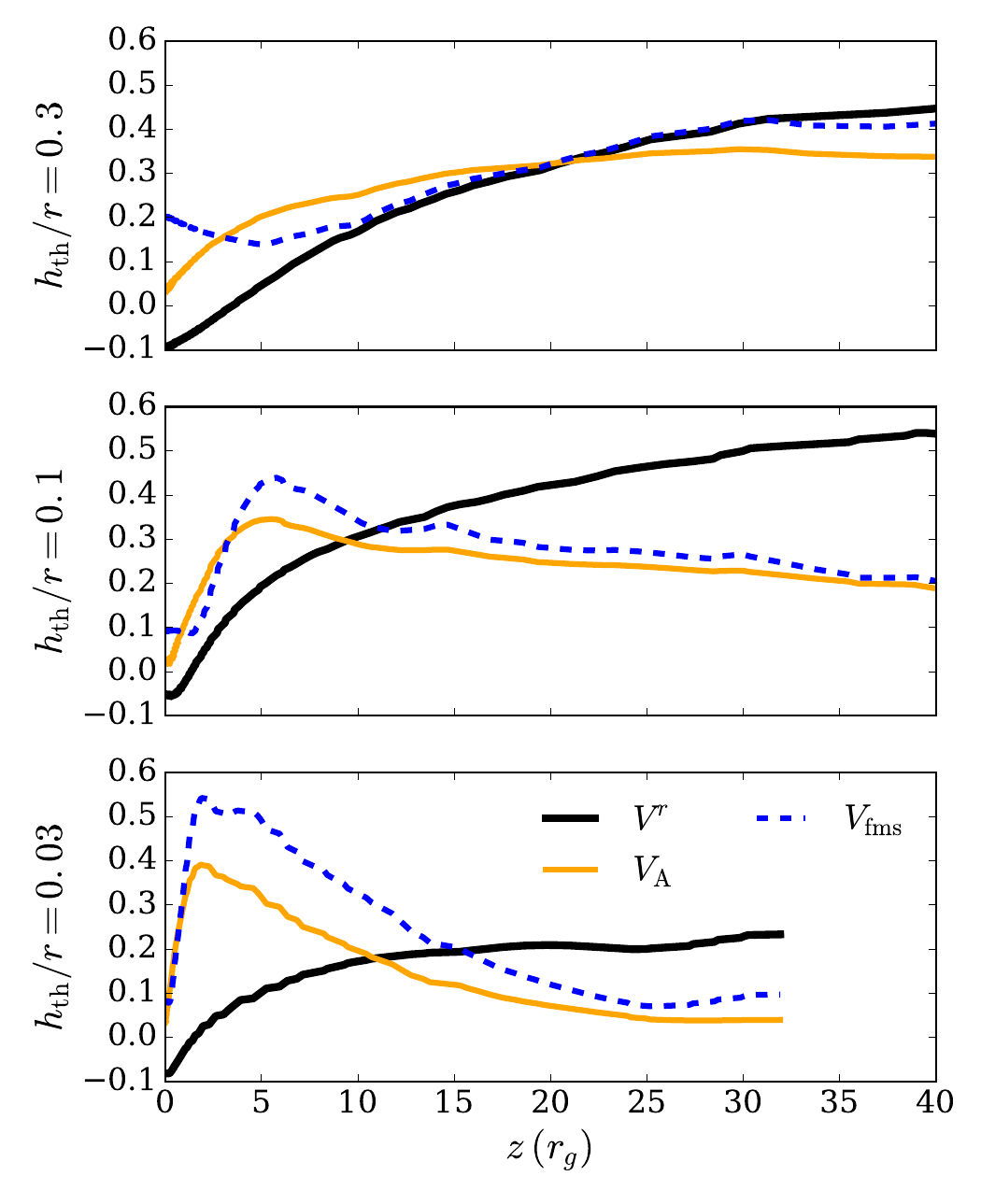}
\caption{Radial velocity (solid black line), Alfv\`en velocity (orange solid line) and fast magnetosonic speed (blue dashed line) as a function of $z$ along a magnetic field line for all $h_\mathrm{th}/r$ in the case where the wind is cooled. The wind crosses all the critical points as it propagates outward.}
\label{fig:wind_velocity}
\end{figure}

Finally, \autoref{fig:wind_velocity} shows the wind velocity as a function of $z$ along a magnetic field field line for each $h_\mathrm{th}/r$. We see that in each case the wind accelerates when propagating outwards and that the velocity seems to saturate at large $z$ to a value between $0.2\:c$ and $0.5\:c$. The field lines we show here crossed the midplane around $3-4\:r_g$ so that the terminal velocity of the wind is of the order of the local launching Keplerian speed. We also see that the wind eventually crosses all the critical points, including the fast magnetosonic point. This makes us confident that the properties of the wind that we reported above are converged and are not the properties of a ``failed wind.''

\section{Discussion}\label{sec:discussion}
\subsection{Long-term evolution and initial conditions}
In this work, we presented the results of MAD simulations with different thermal scale heights,  $h_\mathrm{th}/r$. We used a relatively short duration for our simulations, with a s duration of $44,000\:r_g/c$. We believe that our conclusions drawn from the analysis made for $r<10\:r_g$ should not change much at later times, based on the similarities between our results and longer duration simulations in the literature \citep{avara2016,liska2022,begelman2022}. We tried to to run our thinnest simulation for a longer time but observed a very peculiar behavior in the simulation. Outside of $10\:r_g$ the midplane of the disc starts outflowing and a bump of large density starts propagating inwards. The outflowing material then drags the field lines outwards and the large density bump propagates inwards at the same time. Both these effects destroy our MAD state by the time the simulation has reached $60,000\:r_g/c$. 

\begin{figure*}
\includegraphics[trim={0 10mm 0 0},width=\textwidth]
{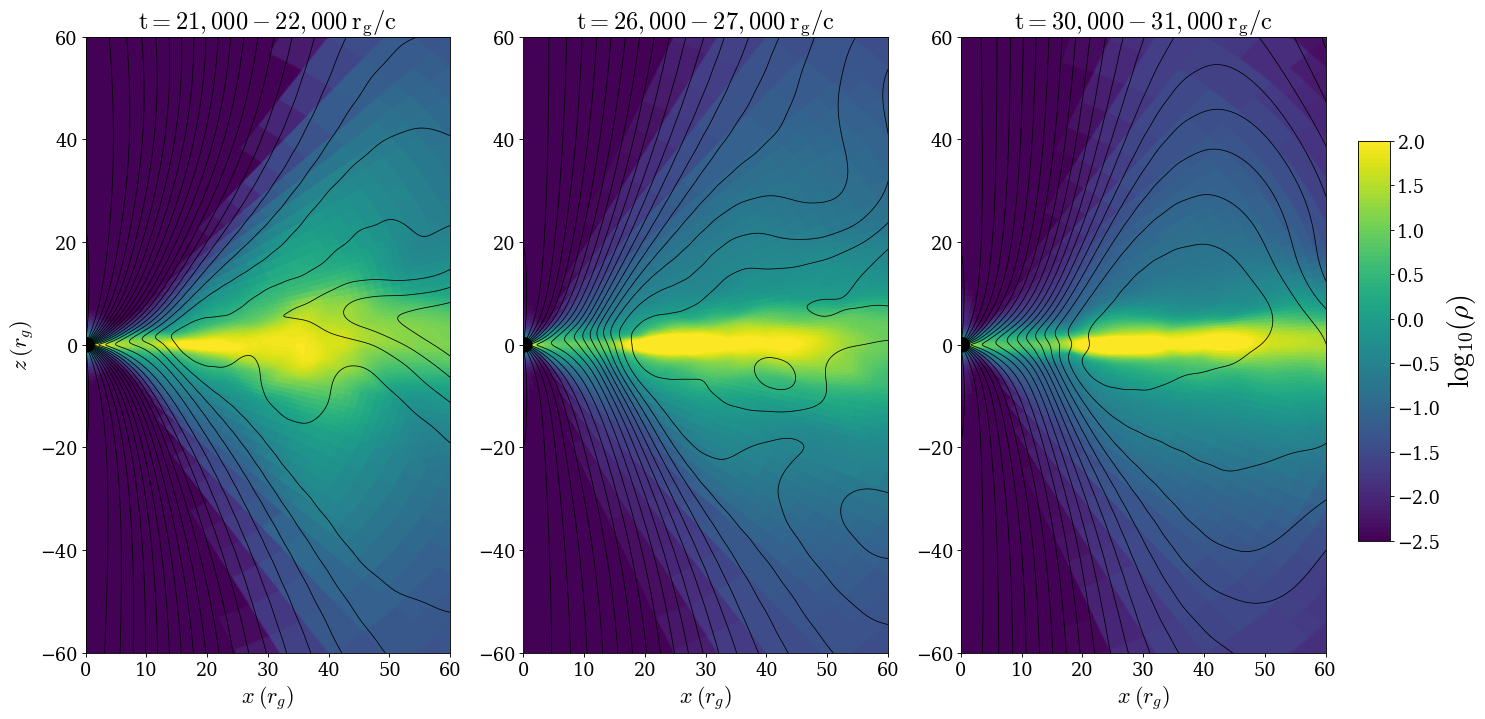}
\caption{Time and $\phi$ averaged maps of the density with the poloidal magnetic field lines superimposed showing the formation of a magnetic loop when we cool down our thick torus with $h_\mathrm{th}/r=0.3$ to a thin disc with $h_\mathrm{th}/r=0.03$. The propagation of this loop inwards ultimately destroys our MAD state in the entire disc leading to the end of the simulation.}
\label{fig:loop_creation}
\end{figure*}

The outflowing material and the density bump seem to be related to the propagation inwards of a large scale poloidal magnetic field loop that was created during the cooling of the torus to a thin disc at $20,000\:r_g/c$ as can be seen on Figure \ref{fig:loop_creation}. We do not see such a large-scale loop appear in our other, hotter simulations. Consequently, we suspect that the creation of this loop is due to the extreme magnetization of our coldest disc. Indeed, in our coldest simulation the pressure exerted by the laminar magnetic structure is larger than the thermal pressure (see \S\ref{fig:force_theta}) and is largely balanced by the turbulent magnetic pressure. However, in the outer disc, which evolve more slowly, the turbulence might not have fully developed. This would allow the magnetic structure to disrupt the initial disc structure and reconnect on itself to create a magnetic loop. Another likely explanation is that the decrease of the refinement level at $r>40\:r_g$ in the outer disc would prevent turbulence to develop and enhance numerical resistiviy so as to promote the creation of this loop. How and when these large-scale loops are created and are they an artifact of our simulation or a physical reality would be an interesting line of research in the future. We note that the outflowing midplane was also observed in \citealt{liska2022} (see negative $\alpha_\mathrm{eff}$ in their Figure 5) and so it is unclear if this is an artifact of thin disc simulations initialized with a torus that is then cooled or if it is a physical effect that would naturally lead to the destruction of thin MADs. In any case, longer duration simulations will be needed to confirm our results and study the behavior of MADs at larger radii. Longer duration simulations will also allow a longer time averaging; this will allow us to reduce noise in our analysis and strengthen our conclusions and possibly remove the asymmetries across the midplane that can be observed in several latitudinal profiles shown here. We emphasize that even if our time window of analysis is relatively short ($4000\:r_g/c$) for the simulation with $h_\mathrm{th}/r=0.03$, the relatively high accretion speed in the simulation allows us to resolve eight inflowing times at $r=10\:r_g$ so that we feel confident we are studying a steady-state up to $10\:r_g$. We also tried widening our time window by time-averaging from $36, 000$ to $44, 000\:r_g/c$. Our main conclusions are unchanged by this choice but, because our disc had an increase in $\dot{M}$ by a factor of 2 between $35, 000\:r_g/c$ and $40, 000\:r_g/c$ that increased the temperature of the disc, the time-averaged thermal scale height is $\approx0.06$ in this time window instead of the targeted value of 0.03.  This is why we did not chose this extended time-window for our main analysis where we focused on the properties of the disc once $\dot{M}$ has reached a steady value (in time and radius up to 10 rg) and the disc has reached its targeted thermal scale height.

We used the same initial conditions for all of our simulations. The fact that the midplane magnetic flux saturates at the same level for all of our simulations could then be an artifact of our simulations starting with the same magnetic flux profile. However, we note that the long-duration simulation in \cite{begelman2022} saturated to a midplane magnetic flux profile that is proportional to radius, close to what is reported in this work, although it started with different initial conditions. Our profile of spherical magnetic flux is also very similar to what is reported in \cite{avara2016}, despite the fact that the other authors used a different initial field profile and initial torus or disc properties. Nonetheless, a proper study of the effect of initial conditions on the saturation profile of MADs and on the mechanisms of flux saturation in MADs remains to be done. It is particularly intriguing that the spherical magnetic flux on the black hole seems to depend on $h_\mathrm{th}/r$ while the midplane magnetic flux  (i.e., the flux threading the disc as a function of radius) does not. If true, this means that a general mechanism of saturation is needed that does not depend on the thermal content of the disc.

\subsection{Convergence of our simulations}\label{sec:convergence}

To test the convergence of our simulations, we first compute the latitudinal MRI quality factor, $Q_\mathrm{MRI,\theta}$, as in \cite{noble2010} but only taken in the midplane to check the convergence of the disc only. We find that we have very good quality factors\footnote{Note that the relevance of the quality factors to deducing the presence of MRI in MADs is not clear \citep{begelman2022}.} of $Q_\theta \approx 22$, $16$ and $16$ for $h_\mathrm{th}/r=0.3$, $0.1$ and $0.03$, respectively, at $r=7r_g$. We note that such high quality factors are a consequence of the very high magnetization in our simulations (see the bottom panel of \autoref{fig:phi_beta}). 

Despite our good MRI quality factors, we note that we only have $\approx5$ cells per $h_\mathrm{th}/r$ for our thinnest simulation with $h_\mathrm{th}/r=0.03$. This could cast doubt on the pertinence of our results, especially since the effects of turbulent magnetic pressure on the disc and the relative importance of large scale stresses compared to turbulent stresses are key results of our paper. To test further the convergence of this simulation, we also compute the latitudinal $Q_\mathrm{corr,\theta}$ and the azimuthal $Q_\mathrm{corr,\phi}$, turbulent quality factors on the turbulent correlation length of the density \citep{shiokawa2012} when $h_\mathrm{th}/r=0.03$. We define 
\begin{equation}
Q_\mathrm{corr,q} = \frac{\lambda_\mathrm{corr,q}-q_0}{\Delta q},
\end{equation}
where $q=\theta$ or $\phi$, $\Delta q$ is the grid cell size in the $q$ direction and $q_0=\theta_\mathrm{min}$ for $q=\theta$ and $q_0=0$ for $q=\phi$. We define $\lambda_\mathrm{corr,q}$ through the following equation
\begin{equation}
\langle R \rangle_t (r,\lambda_\mathrm{corr,q})= \langle {R}\rangle_t (r,0)/e.
\end{equation}
Here $\langle R\rangle_t (r,q)$ is the time-average of the function $R(r,t,q)$, which is the autocorrelation function of $q$ defined for $q=\phi$ as
\begin{equation}
R(r,\phi)= \int_0^{2\pi} \delta q(r,\theta=\pi/2,\phi'+\phi) \delta q(r,\theta=\pi/2,\phi') d\phi' 
\end{equation}
and for $q=\theta$ as
\begin{equation}
R(r,\theta)= \int_0^{2\pi} \int^{\theta_\mathrm{max}}_{\theta_\mathrm{min}} \delta q(r,\theta'+\theta,\phi') \delta q(r,\theta',\phi') d\phi'.
\end{equation}
In what precedes, $\delta q (r,\theta,\phi) = q (r,\theta,\phi)- \langle q \rangle_\phi(r,\theta)$. Finally, we use $\theta_\mathrm{min}=\pi/2-0.3$ and $\theta_\mathrm{max}=\pi/2+0.3$ to confine our analysis to the disc when $q=\theta$. Note that our results are not sensitive to our choice of the $\theta$-extent since they are almost indistinguishable when using a $\theta$-extent going from $\pm0.2$ to $\pm \pi/2$.

We find that for our simulations with $h_\mathrm{th}/r=0.03$ the turbulent quality factors averaged between the horizon and $10\:r_g$ are $Q_\mathrm{corr,\phi}\approx 45$ and $Q_\mathrm{corr,\theta}\approx 6$ compared to a requirement of $\approx 5$ according to \cite{mckinney2013}. To check if our simulation is indeed converged (especially in the $\theta$ direction), we restarted our $h_\mathrm{th}/r=0.03$ simulation at 39,000 $r_g/c$ with an extra-level of refinement at $-0.05 \leq z/r \leq 0.05$, obtaining an effective resolution of $2048\times1024\times2048$ and $10$ points per thermal scale height. We ran the simulation for 4,000 $r_g/c$ and analyzed it between 40,000 and 43,000 $r_g/c$. We do not observe any qualitative change in our results (see Appendix \ref{sec:appendixA} for a comparison of our low and high resolution runs). We computed the new turbulent quality factors from this high resolution simulation and find $Q_\mathrm{corr,\theta}\approx 12$. This means that the typical scale of the turbulent fluctuations did not change between our low and high resolution runs suggesting that we were already resolving the relevant scales. This makes us confident that our simulation was well-resolved with an effective resolution of $1024\times512\times1024$. Finally, we note that the turbulent quality factors indicate that the turbulence length scale is of order $h_\mathrm{th}$. This would be in agreement with the value of $\alpha_{\hat{r}\hat{\phi}}\approx 1$ that we found on \autoref{fig:stresses}. Note that \autoref{fig:alpha} also shows that the level of turbulence in our simulations is in agreement with high resolution shearing box simulations \citep{salvesen2016,scepi2018b}. All of our tests thus point to well-resolved dynamics in the flow. 

A last note concerns the dissipation properties. It is possible that our very extended dissipation profile in $\theta$ could be affected by our derefinement of the grid at high and low latitudes near the poles. Due to our limited computational resources, we could not restart our $h_\mathrm{th}/r=0.03$ simulation with increased resolution up to the latitudes $z/r\approx \pm 0.6$, to which the dissipation extends.  Nonetheless, our simulation with $h_\mathrm{th}/r=0.03$ and an extra level of static refinement near the midplane showed an increase of the dissipation rate in the midplane by only $10\%$ compared to the lower resolution one.\footnote{We note that the inner stress at the ISCO also increased by $10\%$, leading us to suspect that it is the coupling of the black hole to the disc/wind system that was actually affected by our resolution change.} Moreover, our simulation with $h_\mathrm{th}/r=0.1$ has the same level of refinement at $-0.6 \leq z/r \leq  0.6$ and yet shows the same behavior as our simulation with $h_\mathrm{th}/r=0.03$ (extended dissipation profile with two additional wings). This suggests that the dissipation at higher latitudes is not entirely due to our coarser resolution away from the midplane, even though increasing the resolution will certainly have a quantitative effect on the dissipation profile.

\subsection{Implications for observations of black hole accretion discs}\label{sec:implications}
Our results have a number of implications for explaining the rich phenomenology observed in discs around black holes. We will discuss separately each of our main results: coronal dissipation, magnetic support, wind-driven accretion, and outflow properties.

\subsubsection{Coronal emission}\label{sec:corona}
We find that a relatively large amount of dissipation ($30\%$) happens below and above the dense body of the disc forming a ``corona'', which is associated with the base of the wind. This coronal dissipation could maintain hot gas in an outflowing ``sandwich'' geometry even at high accretion rates ($>10\%$ Eddington). This might provide a physical model for part of the X-ray to MeV emission seen in high-luminosity XRBs and AGN \citep{kinch2021} or the soft X-ray excess observed in AGN \citep{petrucci2018}. Our results are very similar to the results of \cite{jiang2019} where the authors find the formation of a hot corona with gas temperatures $\approx 10^9$ K in their radiative, strongly magnetized MHD simulations of an AGN accreting near the Eddington limit. Note that our simulations also have much lower densities than standard discs so that the main body of the disc could also be optically thin for sources shining at $<10\%$ Eddington, forming a corona with a truncated disc geometry (see \S\ref{sec:wind_driven}).

We note that since the coronal dissipation happens in a mildly relativistic outflow, we can expect beaming effects to reduce the formation of reflection features in XRBs and AGNs \citep{reynolds1997,beloborodov1999,malzac2001,markoff2004}. A mildly relativistic outflowing corona would also be consistent with \cite{maccarone2003}, who constrains the maximum outflowing velocity of the corona to be $\approx0.2c$ to reproduce the range of luminosities at which the soft-to-hard state transitions happen among different sources. Although the presence of wind-type outflows during the hard state was originally disfavored by X-ray observations \citep{neilsen2009,ponti2012}, recent optical observations have shown that winds are indeed present in the hard state \citep{munoz2019}, reinforcing the base of the wind as a natural choice to explain the coronal emission.

As mentioned in \S\ref{sec:methods}, we excluded the jet when cooling the domain with our artificial cooling function, preventing a measurement of energy dissipation in the jet. This is because the dynamics of the jet is notoriously affected by the density floors and magnetization ceilings. However, it is not excluded that the jet also dissipates energy that could contribute to the hard X-ray emission of XRBs and AGN. This extra dissipation from the jet could steepen the emission profile of the corona (which is currently $\approx r^{-3}$) in better agreement with reflection models for the iron line, which require a very compact corona \citep{fabian2015}. Assessing the dissipation properties of the jet will require more realistic GRMHD simulations in the high magnetization regime as well as realistic models of the particle microphysics that is believed to play a key role in the jet. We note, however, that recent X-ray polarization measurements have favored a sandwich geometry or a disc geometry for the ``corona'' over very compact emission in a ``lampost'' geometry similar to what X-ray emission from a jet would produce \citep{krawczynski2022}.

\subsubsection{Magnetic support}
Magnetic support has been proposed as a way to suppress the thermal instability of radiation pressure-dominated discs accreting near the Eddington rate \citep{begelman2007}. Indeed, most observations of luminous black holes in XRBs or AGN do not show strong variability although these discs are thought to accrete near Eddington. An exception to this is thought to be the so-called ``heartbeat'' oscillations seen in a few objects of different mass \citep{neilsen2011} and that could be attributed to the radiation pressure instability \citep{wu2016}. However, in most cases there is need of a thermally stable model in this range of accretion rates, to explain observations. 

The thermal instability of radiation pressure dominated discs comes from the strong dependence of the radiation pressure, which supports the disc vertically, on gas temperature \citep{pringle1976,piran1978}. If magnetic pressure provides additional vertical support, this would attenuate the temperature dependence and suppress the instability \citep{sadowski2016,morales2018,mishra2022}. Our simulations show that for discs with $h_\mathrm{th}/r \leq 0.1$ the gradient of turbulent magnetic pressure provides substantial (even dominant for $h_\mathrm{th}/r=0.03$) vertical support and so could thermally stabilize the disc. This is in agreement with highly magnetized simulations with radiative transfer, which show  thermal stability \citep{sadowski2016,morales2018,jiang2019,mishra2022}. This suggests that observations of luminous discs that do not show signs of strong thermal instability, which are most observations, favor highly magnetized disc models.

It also has been suggested that magnetic support could explain the high-luminosity hard state of XRBs by extending the hot, optically thin
solution that is relevant at low accretion rates to a cold, magnetically supported, optically thin solution at high accretion rates close to Eddington \citep{oda2010}. We find that our hot solution (with $h_\mathrm{th}/r=0.3$), although it is extremely magnetized, is not supported by magnetic pressure but almost entirely by thermal pressure and so should collapse when cooling of the electrons and the coupling between protons and electrons becomes important \citep{rees1982}. This is in line with the recent results of \cite{dexter2021} showing that hot MADs are not able to survive to accretion rates higher than $\dot{M} \lesssim 10^{-2}\:\dot{M}_\mathrm{Edd}$, suggesting that a change of geometry of the hard X-ray emission should happen around $10^{-2}\:L_\mathrm{Edd}$. \cite{dexter2021} were not able to follow the thermal evolution of the disc for long after the collapse due to limitations of their radiative scheme in the optically thick regime. However, they find that the disc seems to collapse to a magnetically dominated state with a density scale height $h_\rho/r\approx 0.1$, where $h_\rho\equiv \int \rho|\theta-\theta_0| d\theta/\int \rho d\theta$ and $\theta_0\equiv \frac{\pi}{2} + \int \rho (\theta-\pi/2) d\theta/\int \rho d\theta$ as in \cite{mckinney2012}. This is consistent with our solution with $h_\mathrm{th}/r=0.03$, which has a moderate density scale height of $h_\rho/r\approx 0.1$ supported by magnetic pressure. Our thin-disc, magnetically dominated solution might thus play a key role in the evolution of XRBs (see also \S\ref{sec:corona} and \S\ref{sec:wind_driven}).

\subsubsection{Wind-driven accretion}\label{sec:wind_driven}
The fact that angular momentum transport is dominated by the large-scale (wind-driven) $\theta\phi$ stress rather than MRI turbulent stress leads to much faster inflow speeds than in standard theory. Coupled to the fact that the scale height of the disc is much larger than in standard theory, the fast inflow speed drastically reduces the density in the disc as observed in our thin disc simulation (see \S\ref{sec:vertical}). This effect has also been observed in the simulations of \cite{jiang2019} and \cite{liska2019}. The combination of fast inflow and magnetic support could also explain why the recent thin MAD simulation of \cite{liska2022}, which included radiative transfer, showed a transition from a standard thin disc to a hot, puffy coronal state at high accretion rate, i.e. the formation of a ``truncated disc''. 

The low densities due to wind-driven accretion have been predicted by the self-similar solutions of \cite{ferreira1995} and proposed as a solution to explain the high hard state of XRBs in the so-called jet-emitting disc models (JED; \citealt{ferreira2006,petrucci2010,marcel2018a}). In JED models, the supersonic speed of the flow allows the gas to be optically thin at high accretion rates approaching Eddington, allowing it to produce hard emission at high luminosities in XRBs \citep{marcel2018a,marcel2018b}. With regard to the supersonic inflow velocity due to wind-driven accretion and the consequently low disc density, our cold MAD simulation resembles a JED model. However, our thin MAD simulation also show two characteristics that are absent in a JED: an additional support of turbulent magnetic pressure in the disc and a BH jet at the inner boundary. The influence of the BH jet is particularly important as it provides a deviation from the JED self-similar model. It compresses the flow in the inner disc (see \autoref{fig:force_theta}) and the ergosphere also provides an additional source of energy from the gravitational one. This should be kept in mind when comparing JED and thin MAD models.

The fast inflow speed due to wind-driven angular momentum transport will also affect the temporal evolution of accretion discs. Indeed, wind-driven accretion could induce changes in luminosity on time scales much shorter than viscous time scales, which are traditionally assumed by standard theory. For example, changing-look AGN evolve on very short time scales of months, whereas the viscous time scale would be closer to thousands of years. Wind-driven transport and magnetic support could explain such rapid evolution \citep{dexter2019,scepi2021}. Wind-driven accretion could also explain the relatively short time scales observed during the decay of XRB eruptions \citep{tetarenko2018}.  Shorter inflow times due to wind-driven accretion have also been invoked in CVs as a self-consistent way to explain the recurrence time scales and amplitudes of the observed eruptions, by lowering the density of the inner disc and so delaying the onset of the ionization instability that is thought to drive the eruptions of CVs and XRBs \citep{scepi2019,scepi2020}. A low-density region in the inner parts of XRBs is  needed in any case to reproduce their long recurrence times, and wind-driven accretion could produce such a low-density region instead of or in addition to the evaporation of the inner disc by the hot inner flow \citep{lasota1996,menou2000,dubus2001}.

\subsubsection{Outflows in AGN and XRBs}
The outflows observed in our simulations are reminiscent of the Ultra-Fast Outflows (UFOs) that have been detected in radio-loud and radio quiet AGN \citep{tombesi2010,tombesi2014}. These outflows have velocities of $\approx0.01-0.3c$, are highly ionized and have large mass loss rates estimated to be $\gtrsim 0.05$-$0.1\dot{M}$ and possibly as high as $\dot{M}$ \citep{tombesi2012}. UFOs are suspected to come from the inner regions of AGN \citep{tombesi2012} and so could well be within the reach of our simulations. Indeed, our thin disc simulation exhibits wind velocities very close to the maximum detected in UFOs, with terminal velocities $\approx0.2c$ for the part of the wind originating from $3-4\:r_g$. Unfortunately, we do not have a large enough dynamic range in our simulations to estimate the terminal velocities of outflows originating from $100\:r_g$. However, if the terminal velocity scales as the Keplerian velocity we would estimate a terminal speed of $0.05\:c$ for an outflow originating from $100\:r_g$, in agreement with the range of velocities observed. We can also estimate from our simulations the mass loss in the winds and compare it to UFOs. From our simulations with $h_\mathrm{th}/r=0.03$, for example, we know that the mass loss per log radius is $\approx 0.1\dot{M}$ and so by assuming that the spread in velocities reflects a spread in launching radius we can integrate our local mass loss rate from $3$ to $100\:r_g$ and find a total mass loss of $\approx 1.5\dot{M}$, in agreement with what has been inferred by \cite{tombesi2012}. Semi-analytic MHD models of winds have already been proposed to explain UFOs \citep{fukumura2015} and our simulations demonstrate self-consistently that winds from thin discs can indeed reproduce the observed properties of UFOs without any radiative driving. We also note that \cite{tombesi2013} suggested that the warm absorbers seen in AGN with lower outflow velocities and lower densities might be analogs of the UFOs launched at larger radii. We would need further dynamic range in our simulation to test this hypothesis.

Given that our simulations are scale-free, they would also predict that XRBs have fast outflows similar to AGN if UFOs are to be explained by magnetic driving as we suggest. However, it is very likely that these XRB outflows are too ionized to be detected easily \citep{chakravorty2016}, explaining the lack of X-ray absorption line detections in the hard state (although see \cite{munoz2019}).

\section{Conclusions}\label{sec:conclusion}
In this paper, we present a systematic study of the dynamical properties of magnetically arrested discs for thermal scale heights spanning the range from hot, thick discs to cold, thin discs. We particularly examine the effects of magnetic fields on the radial and vertical structure, angular momentum transport, dissipation, wind and jet properties, and, more generally, deviations from the standard accretion disc theory. We summarize our results as follows: 
\begin{itemize}
    \item The saturation of the magnetic flux on the black hole, a characteristic of MADs, increases with $h_\mathrm{th}/r$ in the disc but, as importantly, with the temperature (and the power) of the wind. This points to a competition between the wind+accretion disc pressure and the jet pressure as the mechanism of saturation of the magnetic flux on the black hole.
    \item The radial magnetic field never supports the disc radially (except maybe in the few inner radii): hot MADs are sub-Keplerian but are mostly supported by the centrifugal force and the thermal pressure while thin MADs are very nearly Keplerian and so are almost entirely supported radially by the centrifugal force.
    \item As discs get colder, the vertical structure of the disc becomes dominated by the magnetic field. For $h_\mathrm{th}/r=0.03$, the density structure is much thicker than expected from standard theory because of vertical support from the turbulent magnetic pressure near the midplane.
    \item In MADs, angular momentum transport is mostly driven by the laminar $\theta\phi$ stress, sometimes called the wind-driven stress, at all $h_\mathrm{th}/r$ studied here. Although the MRI is active and the $r\phi$ stress is large, it has a subdominant role in the overall accretion process because its contribution to the divergence of the stress is less than that of the $\theta\phi$ component. This leads to much faster accretion than in standard theory and thus lower densities in the disc.
    \item In thin MADs, a large part of the local energy  dissipation occurs outside the thermal core ($z/r = \pm 0.03$ for our thinnest model) and a significant fraction (as much as $30\%$) happens in the ouflowing wind, feeding on the latitudinal gradient of angular velocity. Coupled to the fact that the density in the disc is lower because of wind-driven accretion and magnetic support, this promotes dissipation of energy in low-density regions compared to standard theory.
    \item In thin MADs, the disc is more radiatively efficient than predicted by \cite{novikov1973} because of additional stresses inside the ISCO. However, the disc is less radiatively efficient than predicted by \cite{agol2000}, an extension of \cite{novikov1973} taking into account the additional stress inside the ISCO, because of wind-driven accretion.
    \item Fast outflows with terminal velocities of $\approx0.2$ to $0.5c$ and mass loss rates per log radius of $0.1$ to $1\dot{M}$ are found to originate from the inner disc for $h_\mathrm{th}/r=0.03$ and $h_\mathrm{th}/r=0.3$, respectively.
\end{itemize}

All of these deviations from standard theory have implications for explaining the properties of accreting black holes, including: 
\begin{itemize}
    \item Support from the turbulent magnetic pressure should stabilize the disc against thermal instability, such as the radiation pressure instability.
    \item Enhanced dissipation in low-density regions could feed the coronae of XRBs and AGN with enough energy to explain the hard X-ray emission in these objects.
    \item Wind-driven accretion and faster inflow will impact the dynamical evolution of accretion discs and could provide a mechanism to explain the rapid evolution of changing-look AGN, and the long recurrence time scales in XRBs.
    \item The fast magnetic outflows could explain the velocities and mass loss rates of the Ultra-Fast Outflows observed in AGN.
\end{itemize}

\section*{Acknowledgements}
We thank the referee for a very thorough report that improved the robustness of our results. NS is indebted to Christopher White for sharing the implementation of the cooling function used in this work. NS is also grateful to Geoffroy Lesur, Guillaume Dubus, Jonatan Jacquemin-Ide, Pierre-Olivier Petrucci and Jonathan Ferreira for useful discussions concerning the results of this paper. We acknowledge financial support from  NASA Astrophysics Theory Program grants NNX16AI40G, NNX17AK55G, 80NSSC20K0527, and 80NSSC22K0826 and an Alfred P. Sloan Research Fellowship (JD). Part of this work also utilized the Summit supercomputer, which is supported by the National Science Foundation (awards ACI-1532235 and ACI-1532236), the University of Colorado Boulder, and Colorado State University. The Summit supercomputer is a joint effort of the University of Colorado Boulder and Colorado State University.

\section*{Data Availability}
The simulation data analyzed in this article will be shared on reasonable request to the corresponding author.




\bibliographystyle{mnras}
\bibliography{biblio} 

\appendix 

\section{}\label{sec:appendixA}

\begin{figure}
\includegraphics[width=90mm]{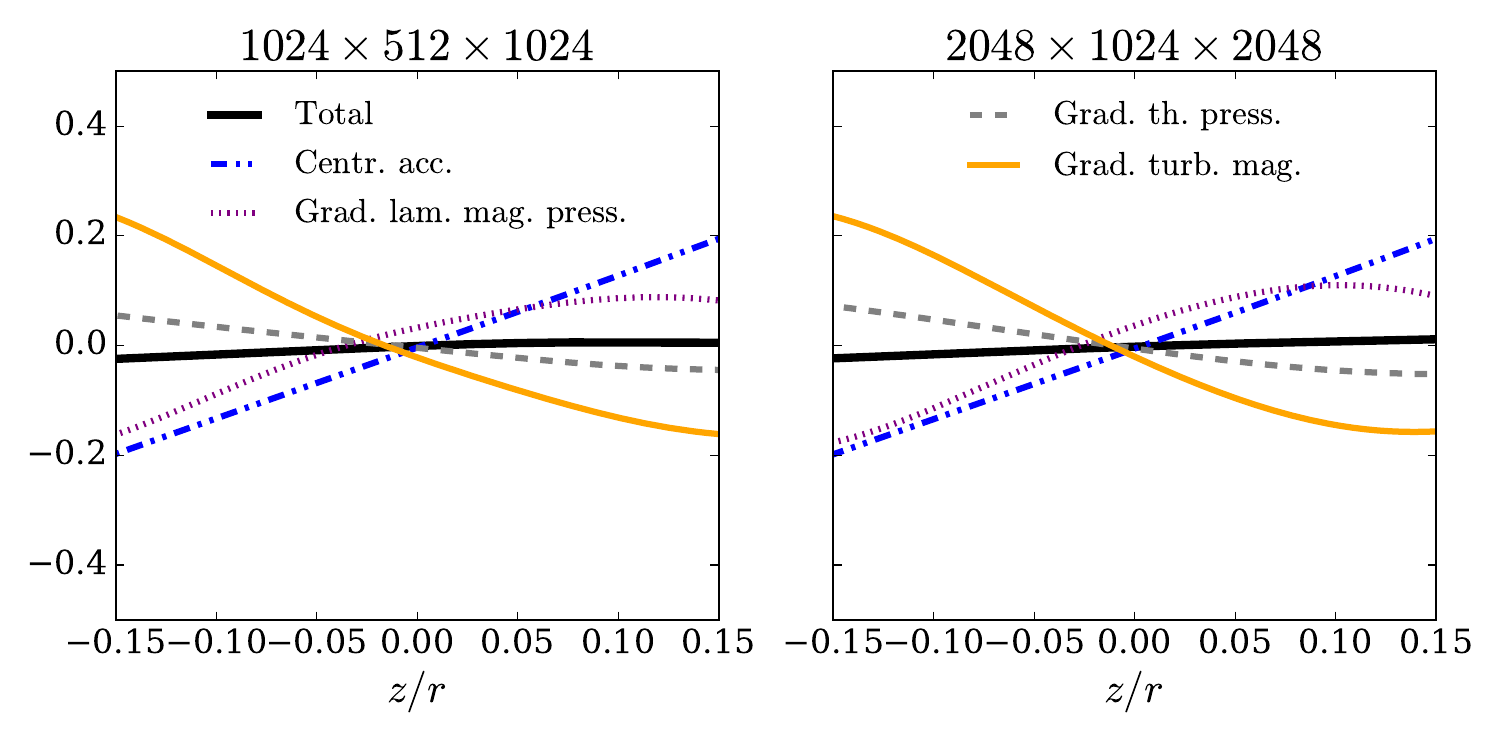}
\caption{Comparison of the latitudinal force budget in our thin disc ($h_\mathrm{th}/r=0.03$) simulations with an effective resolution of $1024\times512\times1024$ (left panel) and of $2048\times1024\times2048$ (right panel). The labelling of the different lines is shared between the left and right panels for graphic purposes. Otherwise, the legend is the same as \autoref{fig:ang_mom_transport}.}
\label{fig:comparison3}
\end{figure}

\begin{figure}
\includegraphics[width=90mm]{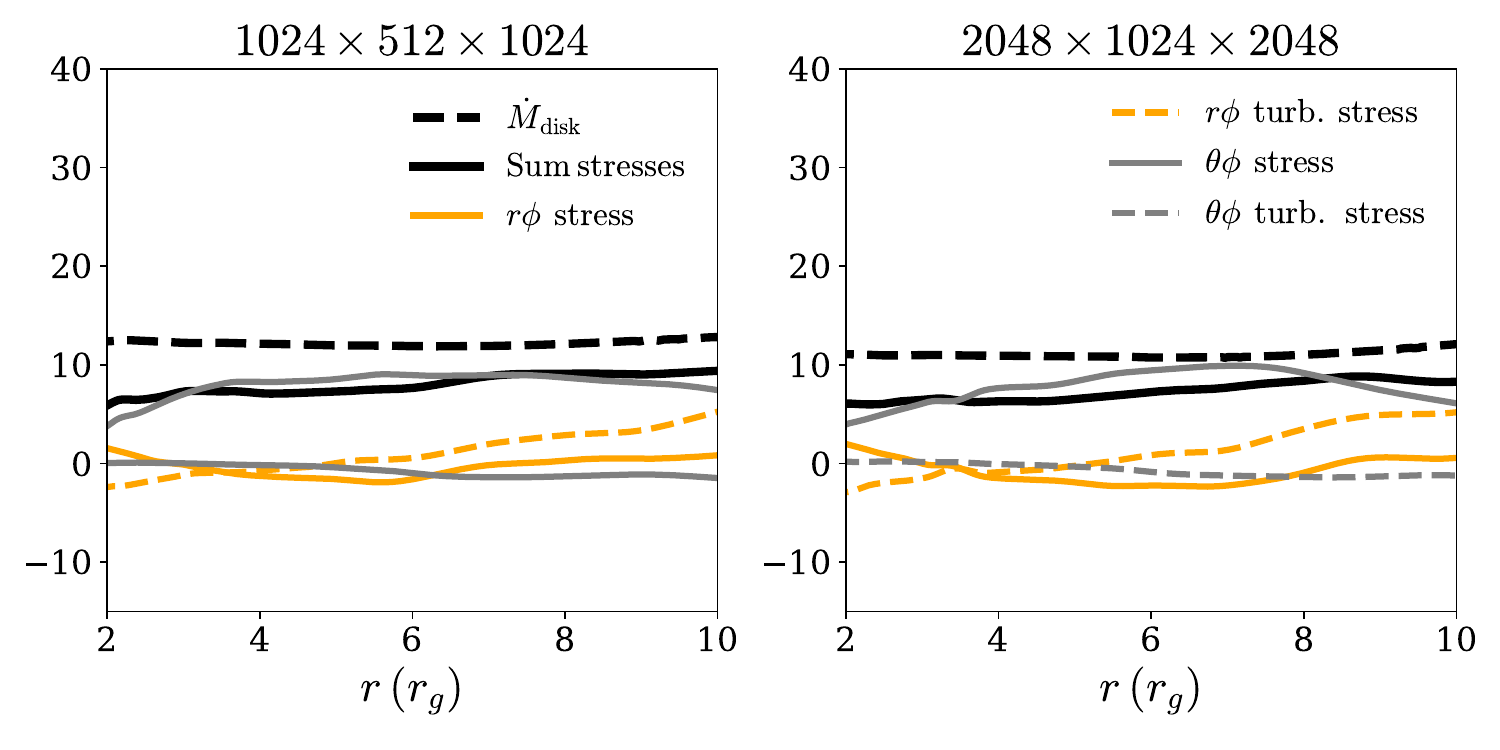}
\caption{Comparison of the contribution of the stress components to the total accretion rate in our thin disc ($h_\mathrm{th}/r=0.03$) simulations with an effective resolution of $1024\times512\times1024$ (left panel) and of $2048\times1024\times2048$ (right panel). The labelling of the different lines is shared between the left and right panels for graphic purposes. Otherwise, the legend of the bottom Figure is the same as \autoref{fig:ang_mom_transport}.}
\label{fig:comparison4}
\end{figure}

\begin{figure}
\includegraphics[width=85mm]{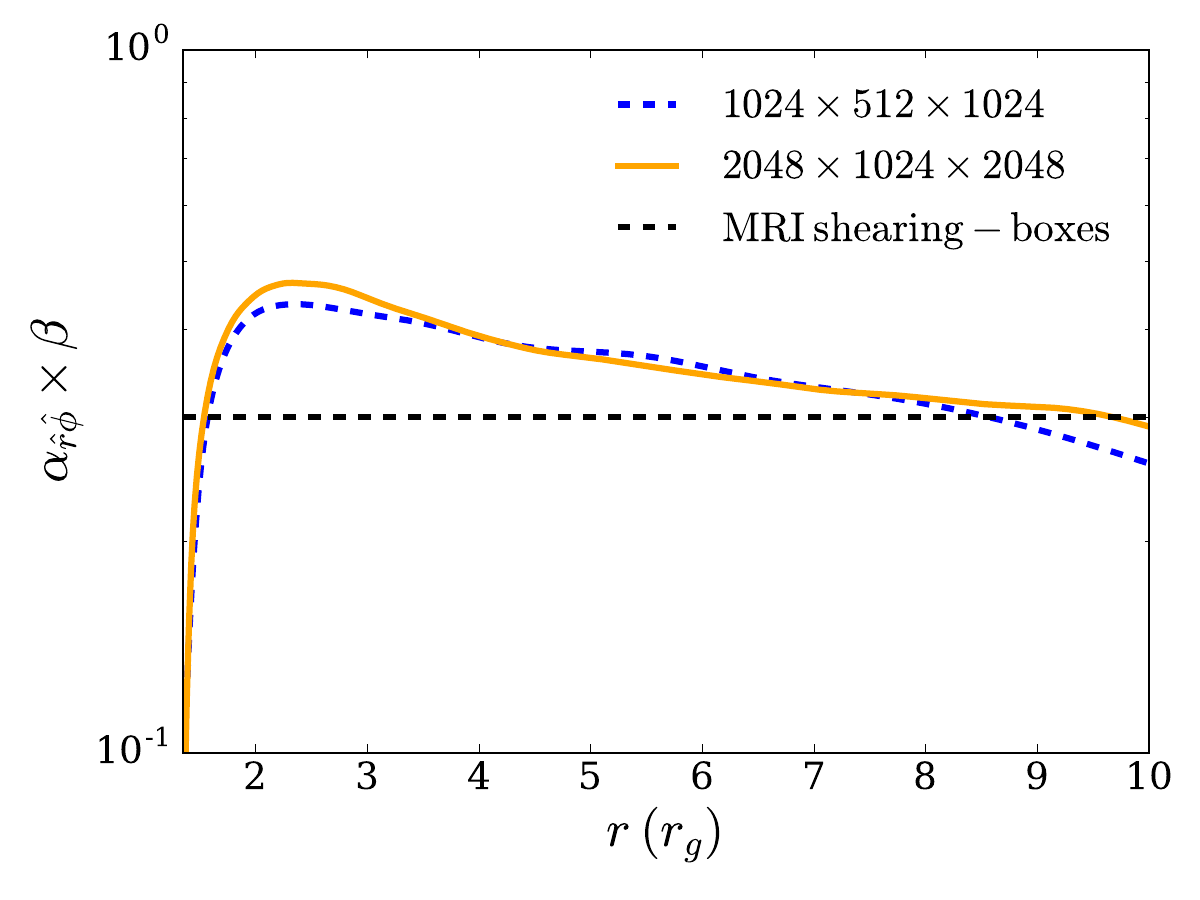}
\caption{Comparison of the measured $\alpha_{\hat{r}\hat{\phi}}$ for our thin disc ($h_\mathrm{th}/r=0.03$) simulations with an effective resolution of $1024\times512\times1024$ (blue dashed line) and of $2048\times1024\times2048$ (solid orange line). The black line shows the expectation from MRI turbulence as in \autoref{fig:alpha}.}
\label{fig:comparison5}
\end{figure}

\begin{figure}
\includegraphics[width=90mm]{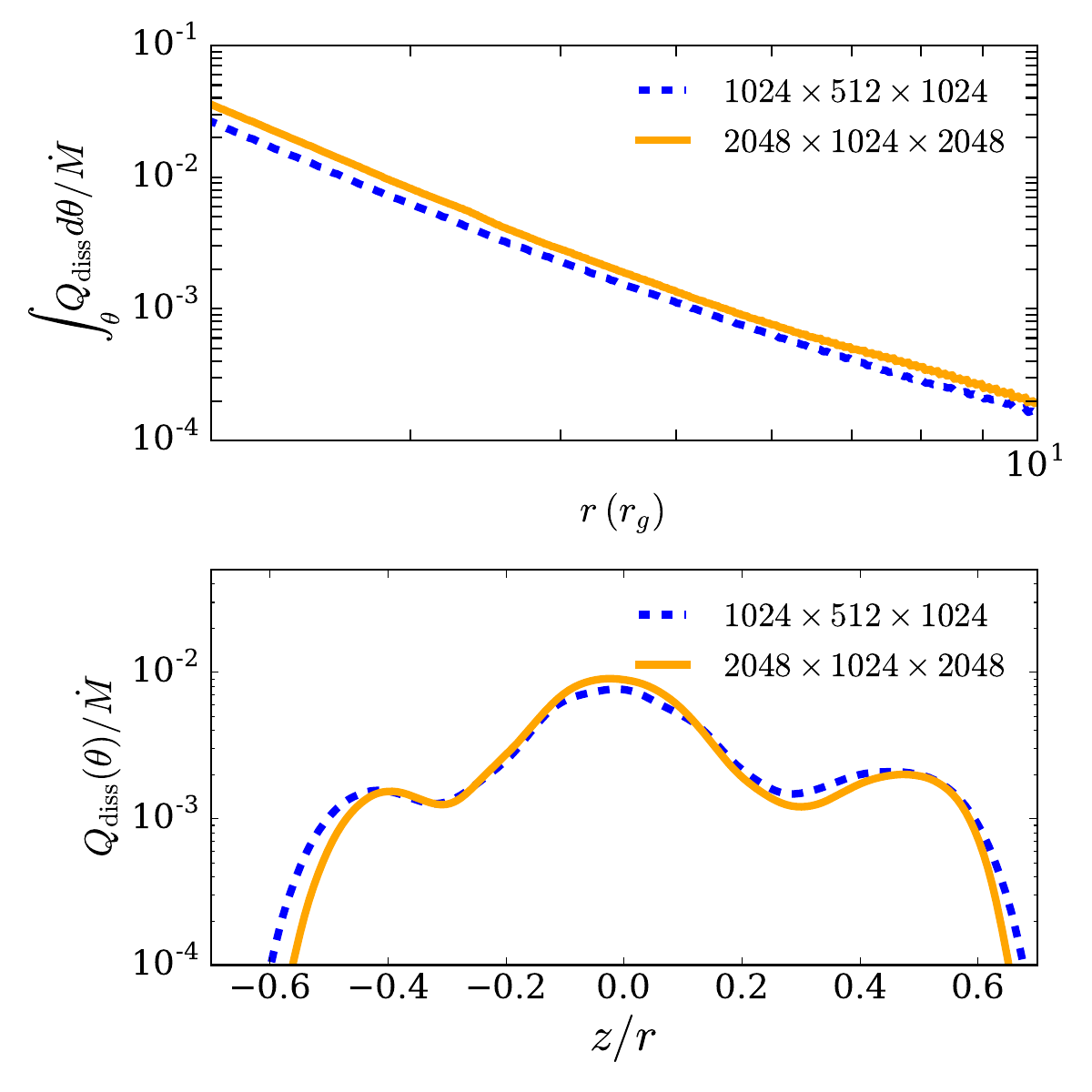}
\caption{Comparison of the radial (top panel) and latitudinal (bottom panel) profile of dissipation normalized by the accretion rate for our thin disc ($h_\mathrm{th}/r=0.03$) simulations with an effective resolution of $1024\times512\times1024$ (dashed blue line) and of $2048\times1024\times2048$ (solid orange line).}
\label{fig:comparison6}
\end{figure}

We plot on \ref{fig:comparison3}, \ref{fig:comparison4}, \ref{fig:comparison5} and \ref{fig:comparison6}, a comparison of our thin disc ($h_\mathrm{th}/r=0.03$) simulations using an effective resolution of  $1024\times512\times1024$ and of $2048\times1024\times2048$. We only show a few plots representative of our main results but we checked that none of the Figures of the paper are greatly affected by the use of a higher resolution. We show here that neither the latitudinal equilibrium force budget (see \autoref{fig:comparison3}), the angular momentum budget (see \autoref{fig:comparison4}), the turbulent $\alpha_{\hat{r}\hat{\phi}}$ (see \autoref{fig:comparison5}), nor the dissipation properties (see \autoref{fig:comparison6}) are greatly affected by the change in resolution, suggesting that the dynamics of our simulation using $1024\times512\times1024$ is well-captured. Slight differences between the low and high resolution simulations can be noticed in \autoref{fig:comparison6}, where we see that the heating rate in the midplane normalized by the accretion rate has increased by $\approx 10\%$ as we increased the resolution. We note that the inner stress at the ISCO (see \S\ref{sec:heating}) has also increased by $\approx 10\%$ meaning that the increase in resolution might have changed by $\approx10\%$ the amount of energy transferred from the black hole to the disc. This is probably due to a better resolved wind/jet interface favoring exchange of energy between the black hole-powered jet and the accretion flow. The increase is not so strong as to imply that we are not resolving the dissipation profile, and so the turbulent properties of the disc correctly. Indeed, the lack of change in the turbulent $\alpha_{\hat{r}\hat{phi}}$ (see \autoref{fig:comparison5}) and in the latitudinal turbulent magnetic pressure support (see \autoref{fig:comparison3}) suggests that we are indeed resolving the turbulence in the core of the disc. Nonetheless, a deeper study of how the disc/black hole energy coupling depends on resolution would be needed in the future.

\bsp	
\label{lastpage}
\end{document}